\documentclass[a4paper,12pt]{article}

\usepackage[english]{babel}
\usepackage{braket}

\usepackage{color}
\usepackage{tikzsymbols}
\usepackage{amsmath}
\usepackage{amssymb}
\usepackage{amsfonts}
\everymath{\displaystyle}
\usepackage{tikz}
\usepackage{physics}
\usetikzlibrary{shapes.misc}
\usetikzlibrary{shadings}

\usepackage{color}
\usepackage{enumerate}

\def\zb{{\overline{z}}}
\def\cD{{\mathcal{D}}}
\def\cO{{\mathcal{O}}}

\definecolor{ws1}{RGB}{217,227,244}
\definecolor{ws2}{RGB}{152,178,222}

\usepackage{amsmath}
 \makeatletter
    
    \@addtoreset{equation}{section}
  \makeatother
\usepackage{graphicx}
\usepackage[colorlinks=true, allcolors=blue]{hyperref}

\begin{document}

\baselineskip=18pt  
\numberwithin{equation}{section}  



\thispagestyle{empty}

\vspace*{-2cm} 
\begin{flushright}
KEK-TH-2758\\
YITP-25-138\\
\end{flushright}

\vspace*{1.5cm} 
\begin{center}
 {\LARGE Black Hole Entropy from String Entanglement}\\
 \vspace*{1.3cm}
 Soichiro Mori$^1$, Tadakatsu Sakai$^{2,1,3}$ and Masaki Shigemori$^{1,4}$\\
 \vspace*{1.0cm} 
 $^1$ 
Department of Physics, Nagoya University\\
Furo-cho, Chikusa-ku, Nagoya 464-8602, Japan\\[1ex]
 
 $^2$ 
Kobayashi-Maskawa Institute for the Origin of Particles and the Universe, \\Nagoya University,\\
Furo-cho, Chikusa-ku, Nagoya 464-8602, Japan\\[1ex]

 $^3$ 
 KEK Theory Center, Institute of Particle and Nuclear Studies,\\ High Energy Accelerator Research Organization,\\ 1-1 Oho, Tsukuba, Ibaraki 305-0801, Japan\\[1ex]

$^4$ 
Center for Gravitational Physics,\\
Yukawa Institute for Theoretical Physics, Kyoto University\\
Kitashirakawa Oiwakecho, Sakyo-ku, Kyoto 606-8502, Japan
\end{center}
\vspace*{0.8cm}

\noindent
We discuss the notion of string entanglement in string theory, which aims to study entanglement between worldsheet Hilbert spaces rather than entanglement between spacetime Hilbert spaces defined on a time slice in spacetime.
Applying this framework to the FZZ duality and its extension to a three-dimensional black hole, we argue that the thermal entropy of 2d and 3d black holes is accounted for by the string entanglement entropy between folded strings arising in the dual sine-Liouville CFT.
We compute this via a worldsheet replica method and show that it decomposes into two parts, which we call the vertex operator contribution and the replica contribution. The former can be evaluated analytically and is shown to coincide with the black hole thermal entropies in the low temperature limit in  large $D
$ dimensions. 
Although a computation of the latter is left as an open problem, we present evidence that it captures the remaining portion of the black hole entropy.

\newpage
\setcounter{page}{1} 



\tableofcontents


\section{Introduction}

The idea to understand black hole entropy as entanglement entropy across the horizon has been around for quite some time \cite{Sorkin:1984kjy,  Bombelli:1986rw} (see also \cite{Witten:2024upt}). 
This idea finds a firmer ground \cite{Maldacena:2001kr} in the context of the AdS/CFT correspondence, which states that an eternal, two-sided AdS black hole is holographically dual to a thermo-field double state for two boundary CFTs, and that the black hole entropy is the entanglement entropy between the two CFTs.   Nevertheless, one may still ask whether black hole entropy admits an interpretation as entanglement entropy between degrees of freedom across the horizon.  In string theory, one naturally wonders if black hole entropy can be reproduced from the entanglement entropy of strings across the horizon~\cite{Susskind:1993ws, Susskind:1994sm}.

\begin{figure}[htbp]
\begin{center}
\includegraphics[height=40mm]{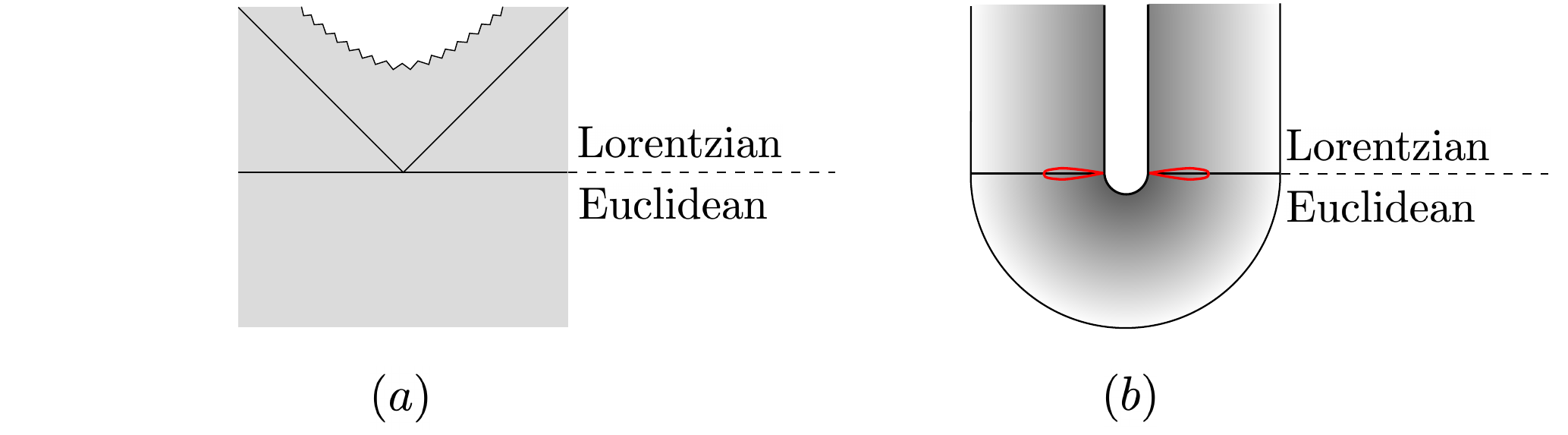}
 \caption{\it 
 FZZ duality and stringy ER=EPR\@. The FZZ duality relates (a) the cigar CFT background and (b) the sine-Liouville background, where the gradient depicts the dilaton profile.  Upon continuation from Euclidean signature to Lorentzian signature, in (a), the Euclidean cigar half-disk is continued to a Lorentzian black hole in which two asymptotic regions are connected by an Einstein-Rosen (ER) bridge, while in (b) the Euclidean section, which is topologically is a half-annulus, is continued to two disconnected flat spacetimes. The ER bridge in (a) corresponds to the EPR pairs, which are realized as pairs of open folded strings as shown in red in (b).
 }  \label{FZZ duality and string ER=EPR}
\end{center}
\end{figure}

An interesting observation along this line was made
recently by Jafferis and Schneider in~\cite{Jafferis:2021ywg}.
The Fateev-Zamolodchikov-Zamolodchikov (FZZ) duality  \cite{FateevZamolodchikovUnpublished,Kazakov:2000pm} is a duality that relates the two-dimensional cigar CFT \cite{Witten:1991yr} and the sine-Liouville CFT\@.  The cigar CFT describes string propagation in a two-dimensional Euclidean black hole background. As in any black hole backgrounds, the thermal circle shrinks to zero on the horizon, which is thought to be a result of condensation of strings that winds around the thermal circle, although no thermal string is present explicitly in the CFT action.\footnote{It is argued that condensation of folded strings occurs inside the cigar black hole \cite{Itzhaki:2018glf}.}
In the dual sine-Liouville CFT, on the other hand, there is no horizon but instead there is an explicit condensate of thermal strings in the action.
The observation of \cite{Jafferis:2021ywg} is that,
upon continuation to Lorentzian signature, the cigar side becomes a two-sided black hole in the Hartle-Hawking state, while the sine-Liouville side involves a condensate of entangled open folded strings living in two separate spacetimes.  This can be regarded as 
a string version of the ER=EPR conjecture \cite{Maldacena:2013xja}, in which the Einstein-Rosen bridge of the two-dimensional Lorentzian black hole corresponds to the entangled pairs of strings on the sine-Liouville side \cite{Jafferis:2021ywg}.  See Figure \ref{FZZ duality and string ER=EPR}.

It is then natural to ask whether the black hole entropy of the cigar black hole can be interpreted as the entanglement entropy between the entangled pairs of strings on the sine-Liouville side.  
We refer to this as string entanglement entropy, because it is interpreted as a measure of entanglement among worldsheet Hilbert spaces.
In this paper, we compute the string entanglement entropy using a worldsheet replica trick on the sine-Liouville side, and find that it reproduces the black hole entropy in the classical limit.  More precisely, we find that the string entanglement entropy consists of two contributions: the ``vertex operator contribution'' and the ``replica contribution''.  We explicitly evaluate the former and find that it reproduces the known thermal entropy of the cigar black hole in the classical limit.
Assuming that the total entanglement entropy agrees with
the expectation from the thermal entropy of the cigar black hole including $\alpha'$ corrections computed in \cite{Halder:2024gwe,Callan:1988hs}, we can predict the other, replica contribution to the entanglement entropy.
Moreover, we extend our result to the entropy of the BTZ black hole using the 3d version of the FZZ duality
\cite{Halder:2022ykw}.
It is interesting to note that the 2d and 3d results in the vertex operator contribution exhibit a universal behavior that they take the same form except the partition function of an internal CFT\@.
We leave it for a future work to derive the replica contribution.

The structure of the rest of the paper is as follows.
In section \ref{sec:FZZ_duality_and_string_EE}, we give a brief review of the FZZ duality with an emphasis on how folded strings emerge from condensation of winding string vertex operators. In section \ref{sec:string_entanglement_in_sL_string}, we employ a worldsheet replica method to compute the string entanglement entropy associated with folded strings, and then compare the result with the $\alpha^\prime$-corrected black-hole thermal entropy in large $D$ dimensions. An extension to three-dimensional black hole is also made in this section. 
We conclude this paper with summary and outlook.
In Appendix \ref{app:replica_with_vtx}, some technical details about the worldsheet replica method in the sine-Liouville CFT are demonstrated.
In Appendix \ref{app:String theory with non-marginal vertex operators}, we discuss how to evaluate correlation functions in the Polyakov string on a Riemann sphere with non-marginal vertex operators inserted.

\section{FZZ Duality and String Entanglement Entropy}
\label{sec:FZZ_duality_and_string_EE}

In this section, we argue that 
Fateev-Zamolodchikov-Zamolodchikov (FZZ) duality
provides us with a natural framework for studying string
entanglement entropy, based on the work by Jafferis and 
Schneider\,\cite{Jafferis:2021ywg}. They showed that
the FZZ duality can be interpreted as a manifestation of the ER=EPR 
correspondence\,\cite{Maldacena:2013xja} (see also \cite{VanRaamsdonk:2010pw}), 
which relates a disconnected spacetime with entangled degrees of freedom 
on it to a connected spacetime with an Einstein-Rosen 
bridge. 
The FZZ duality is a duality between the cigar CFT and the sine-Liouville
CFT, originally conjectured by Fateev, Zamolodchikov, and Zamolodchikov 
\cite{FateevZamolodchikovUnpublished}, further studied by Kazakov, Kostov, and
Kutasov \cite{Kazakov:2000pm}, and proven by Hikida and 
Schomerus \cite{Hikida:2008pe}.
A key ingredient in \cite{Jafferis:2021ywg} is to make Lorentzian continuation
of the two CFTs appearing in the FZZ duality; the cigar CFT
is continued to that on a two-sided, Lorentzian black hole 
while the target space of the sine-Liouville CFT is continued to two
disconnected spacetimes with the time direction specified by a 
Schwinger-Keldysh contour.
As first found in \cite{Jafferis:2021ywg} and reviewed shortly in this paper,
the entangled degrees of freedom in the disconnected spacetime
are given by folded strings in the sine-Liouville CFT.
We are thus led to an idea that the thermal black hole entropy
can be accounted for by the entanglement entropy of the folded
strings.
In this paper, we examine if this idea is true or not
by evaluating the string entanglement entropy associated with the
folded strings by means of a replica trick.
For this purpose, this section starts by giving a brief review 
of the FZZ duality and then discusses how folded strings arise
by studying the sine-Liouville partition function, following
the results of \cite{Jafferis:2021ywg}.

Before proceeding, some comments are in order.
We use the term the ``sine-Liouville string'' 
to mean a string theory consisting of the sine-Liouville CFT sector, 
the internal CFT sector and the ghost CFT sector with the total central charge
equal to zero. 
Similarly, we use the term the ``cigar string'' to mean a string theory 
consisting of the cigar CFT sector, the internal CFT sector and 
the ghost CFT sector
with the total central charge equal to zero.

\subsection{Review of the FZZ duality}\label{Review of the FZZ Duality}

{}For the purpose of constructing 
the cigar CFT, we start from the \( SL(2,\mathbb{C})_k/SU(2) \) 
gauged Wess-Zumino-Witten (WZW) model \cite{Witten:1991yr,Jafferis:2021ywg}. 
Here, \( SL(2,\mathbb{C})/SU(2) \) is regarded as the Euclidean 
BTZ black hole with the topology of a solid torus.\footnote{More precisely, \( SL(2,\mathbb{C})/SU(2) \) is the universal cover of the Euclidean BTZ black hole and we must divide it by $\mathbb{Z}$.}
The cigar CFT is obtained by gauging a $U(1)$ isometry of the solid 
torus acting on a non-vanishing 1-cycle corresponding to the event horizon of the BTZ black hole and integrating the associated gauge 
field. We are then 
led to the \( SL(2,\mathbb{C})_k/(SU(2) \times U(1)) \) CFT\@. 
The Virasoro central charge of the CFT is given by 
\begin{align}
c=\frac{3k}{k-2}-1 \ .    
\label{c:cigar}
\end{align}
In the large $k$ limit, the action reads
\begin{align}
    S_{\mathrm{cigar}}=\frac{k}{4\pi}\int_\Sigma d^2\sigma\sqrt{h}\left((\nabla r)^2+\tanh^2r(\nabla\theta)^2+\frac{1}{k}R[h](\Phi_0-\ln\cosh r)\right)\ ,\label{cigar_CFT_action}
\end{align}
where \( r \in [0, \infty) \) and \( \theta \in [0, 2\pi) \) are 
the coordinates of the target space, \( \Sigma \) represents 
the worldsheet, \( h \) is the worldsheet metric, and \( R[h] \) denotes 
the worldsheet Ricci curvature. The target space is a two-dimensional ``cigar'' with the radius of $S^1_\theta$ smoothly shrinks to zero at $r=0$ while it asymptotes to constant as $r\to \infty$.
The winding number conservation law
is broken because the $S^1$ is not a nontrivial cycle, shrinking to zero at the tip of the cigar.
Note that $k>2$ is required for $c>0$ and for physical quantities in the dual sine-Liouville CFT to be well-defined. In this paper we assume $k> 3$ so that we can sensibly interpret the gauged WZW model as a string theory on the two-dimensional black hole. This is because the winding scalar field $\hat{\chi}$ on the cigar background behaves as $\hat{\chi} \sim e^{-(k-3)r}$ for $k>3$ as $r\to\infty$ \cite{Kazakov:2000pm}. This is also understood by the fact that the effective mass  squared of the winding scalar is proportional to $(k-3)^2$ due to the contribution from a linear dilaton gradient for $r\to\infty$ \cite{Karczmarek:2004bw, Maldacena:2005hi,Chen:2021emg}. See \eqref{Phi_cigar} for the dilaton profile in the cigar CFT\@. It follows that the winding scalar field becomes non-normalizable unless $k>3$; otherwise the contribution of $\hat{\chi}$ to the black hole thermal entropy diverges. The partition function of the cigar CFT is
\begin{align}
    Z_\mathrm{cigar}=\int\mathcal{D}r\, \mathcal{D}\theta\, e^{-S_{\mathrm{cigar}}}.\label{cigar-partition-function}
\end{align}

The sine-Liouville CFT is composed of a linear dilaton CFT and a 
compact free boson ($\mathrm{LD} \times S^1$) with a potential
term due to a winding string condensate. The action is given by
\begin{align}
    S_{\mathrm{sL}} &= S_{\mathrm{LD} \times S^1} + 2\lambda \int_\Sigma d^2\sigma \sqrt{h} \, V_\mathrm{sL}\ , 
\label{sine-Liouville action}
\end{align}
where 
\begin{align}
    S_{\mathrm{LD} \times S^1} := \frac{1}{4\pi} \int_\Sigma d^2\sigma \sqrt{h} \left( (\nabla \hat{r})^2 + (\nabla \hat{\theta})^2 - Q R[h] \hat{r} \right)
\ .
\end{align}
Here,  $\hat{r} \in (-\infty, \infty)$ and $\hat{\theta} \in [0, 2\pi\sqrt{k})$
denote the target space coordinates. 
The background charge $Q$ is fixed by equating the central charge of $\mathrm{LD} \times S^1$ with \eqref{c:cigar} to be\footnote{In this paper, we work in units of $\alpha^\prime=1$.} $Q = 1/\sqrt{k-2}$.
The potential $V_\mathrm{sL}$ originates from condensation of
string states with a unit winding number and is given by
\begin{align}
    V_\mathrm{sL} := \frac{1}{2}(W_+ + W_-), \quad
    W_\pm := e^{-2b_{\mathrm{sL}} \hat{r}} e^{\pm i \sqrt{k} (\hat{\theta}_\mathrm{L} - \hat{\theta}_\mathrm{R})}\ ,
\label{sine-Liouville potential}
\end{align}
Here, $\hat{\theta}_{L,R}$ denote the left- and right-moving mode
of $\hat{\theta}$, respectively. 
The vertex operator 
$e^{\pm i\sqrt{k}(\hat{\theta}_\mathrm{L} - \hat{\theta}_\mathrm{R})}$
carries a unit winding number along the $S^1$, and is dressed 
by a primary operator \( e^{-2b_\mathrm{sL}\hat{r}} \).
The linear-dilaton momentum $-2b_\mathrm{sL}$ is determined by requiring that 
$W_\pm$ be marginal
\begin{align}
    b_\mathrm{sL}(Q - b_\mathrm{sL}) + \frac{k}{4} = 1 \ .
\end{align}
We find $b_\mathrm{sL} = \tfrac{1}{2} \sqrt{k-2}$.

Finally, we comment on how
the two CFTs behave in asymptotic regions of $r$ and $\hat{r}$. The dilaton coupling to the Ricci scalar in the sine-Liouville CFT action implies that the dilaton is given by 
\begin{align}  
  \exp\Phi_{\rm sL}(\hat r) = g_s  e^{-Q\hat{r}} \ .  \label{Phi_sL}
\end{align}
This diverges as \( \hat{r} \to -\infty \). However, the sine-Liouville potential \eqref{sine-Liouville potential} grows to infinity as well, preventing strings from propagating to the strong coupling region. 
In contrast, the target space geometry in the cigar CFT terminates at the tip $r=0$, where the dilaton  
\begin{align}
    \exp\Phi_{\mathrm{cigar}}(r)= e^{\frac{1}{k}(\Phi_0-\ln\cosh{r})}\label{Phi_cigar}
\end{align}  
is left finite.

In the asymptotic region $\hat{r}\to\infty$, the sine-Liouville CFT reduces to the linear dilaton times the compact free boson, because the sine-Liouville potential \eqref{sine-Liouville potential} vanishes there. In the cigar CFT, the asymptotic behavior for $r\to\infty$  is also described by the linear dilaton CFT times the free boson on \( S^1 \). Therefore, the target spaces described by the two CFTs have the same asymptotic behavior for  $r,\hat{r}\to\infty$.

\subsection{Partition Function of Sine-Liouville CFT}
To see how folded strings arise in sine-Liouville string, let us analyze the partition function of the sine-Liouville CFT
\begin{align}
    \hat{Z}_\mathrm{sL} &= \int \mathcal{D}\hat{r}\ \mathcal{D}\hat{\theta}\ e^{-S_{\mathrm{sL}}}.
\end{align}
Let the constant mode and non-constant mode of $\hat{r}$ be $\hat{r}_0$ and $\hat{r}'$, respectively.  Then, the partition function becomes
\begin{align}
    \hat{Z}_\mathrm{sL} &= 2\pi\int \mathcal{D}\hat{r}'\, \mathcal{D}\hat{\theta}\, e^{-S_{\mathrm{LD\times} S^1}[\hat{r}',\hat{\theta}]}
    \int d\hat{r}_0 \, 
    \exp(\chi Q\hat{r}_0 - \lambda e^{-2b_{\mathrm{sL}}\hat{r}_0} \int d^2\sigma \sqrt{h} \, \big(W_+(\hat{r}',\hat{\theta}) + W_-(\hat{r}',\hat{\theta})\big)),
\end{align}
where $\chi$ is the Euler characteristic of the worldsheet. Here, we focus only on the spherical worldsheet so that $\chi=2$. 
The prefactor $2\pi$ comes from the normalization of the integral measure for $\hat{r}_0$, which is fixed from an analysis of a three-point function in $SL(2,\mathbb{C})_k/U(1)$ coset CFT in \cite{Halder:2022ykw}. Using the following formula \cite{Halder:2023adw}
\begin{align}
    \int_{-\infty}^\infty d\hat{r}_0\ \exp(-2a\hat{r}_0 - \alpha e^{-2b\hat{r}_0}) = \frac{1}{2b}\Gamma\!\left(\frac{a}{b}\right)\alpha^{-\frac{a}{b}},
\end{align}
we find
\begin{align}
    \hat{Z}_\mathrm{sL} &= \frac{2\pi}{b'} \Gamma(-2s') \left(\frac{\mu}{2b'^2}\right)^{2s'} \nonumber \\
    &\quad \times \int \mathcal{D}\hat{r}'\ \mathcal{D}\hat{\theta}\ e^{-S_{\mathrm{LD\times}S^1}[\hat{r}',\hat{\theta}]}
    \left(\int d^2z \, W_+(\hat{r}',\hat{\theta}) + \int d^2z \, W_-(\hat{r}',\hat{\theta})\right)^{2s'},
    \label{Z_sl_W_brought_down}
\end{align}
where we have introduced the parameters \( b' := 2b_{\mathrm{sL}} \), \( \mu := 4\lambda b'^2 \), and 
\begin{align}
 s' := {1\over b'^2} = {1\over k-2}
\end{align}
following \cite{Halder:2024gwe}. $Z_\mathrm{sL}$ can be regarded as a $2s'$-point function of the vertex operators \( W_\pm \) in the $\mathrm{LD\times}S^1$ CFT\@. This is justified by taking $s'$ to be a positive integer for the moment. Eventually,  \( s' \) will be analytically continued to a positive real value after the path integral is performed.

Let \(\hat{\theta}_{0L,R}\) be the constant mode of \(\hat{\theta}_{L,R}\), respectively. The integral measure for the constant mode should be defined in terms of $\hat{\theta}_{0L}-\hat{\theta}_{0R}$, because the inserted vertex operator $W_{\pm}$ is written in terms of $\hat{\theta}_{L}-\hat{\theta}_{R}$. In order to specify the period of $\hat{\theta}_{0L}-\hat{\theta}_{0R}$, we T-dualize the  $\mathrm{LD\times}S^1$ CFT along the $S^1$. Then, $W_{\pm}$ is mapped to a vertex operator of a unit momentum along the T-dualized circle of radius $1/\sqrt{k}$.
As the T-duality transformation should leave the partition function unchanged, the integral over the constant mode $\hat{\theta}_{0L,R}$ must give $2\pi/\sqrt{k}$. Therefore, after carrying out the $\hat{\theta}_{0L,R}$ integral, we obtain
\begin{align}
    \hat{Z}_\mathrm{sL} &= \frac{2\pi}{\sqrt{k}}\cdot \frac{2\pi}{b'} \Gamma(-2s') \left(\frac{\mu}{2b'^2}\right)^{2s'} \frac{\Gamma(2s'+1)}{\Gamma(s'+1)^2} \nonumber \\
    &\quad \times \int \mathcal{D}\hat{r}'\ \mathcal{D}\hat{\theta}'\ e^{-S_{\mathrm{LD\times}S^1}[\hat{r}',\hat{\theta}']} 
    \left(\int d^2z \, W_+(\hat{r}',\hat{\theta}')\right)^{s'} \left(\int d^2z \, W_-(\hat{r}',\hat{\theta}')\right)^{s'}.
\end{align}

$\hat{Z}_\mathrm{sL}$ is ill-defined because the integration over the location of the vertex operators gives rise to an overall divergence. Since $W_{\pm}$ is marginal with conformal dimension $(1,1)$, it is proportional to the volume of $SL(2,\mathbb{C})$ transformation. The divergence is then removed by dividing $\hat{Z}_\mathrm{sL}$ by the  $SL(2,\mathbb{C})$ volume. This is nothing but the Faddeev-Popov procedure in the Polyakov string, where $SL(2,\mathbb{C})$ is identified with the conformal Killing group acting on a string worldsheet with the topology of a sphere. The Jacobian associated with the change of integration measure from the one for the position of three vertex operators to the $SL(2,\mathbb{C})$ Haar measure is equal to the three-point function of the conformal $c$-ghost, which is denoted by $Z_c$. The position of the three vertex operators is fixed to be any reference points. 
See appendix \ref{app:String theory with non-marginal vertex operators} for a review.

Let $Z_\mathrm{sL}$ be the partition function of the sine-Liouville CFT after
removing the $SL(2,\mathbb{C})$ volume from $\hat{Z}_\mathrm{sL}$, where two $W_{+}$ insertions are placed at $z=0$ and
$z=1$, and one $W_{-}$ is placed at $z=\infty$, namely,
\begin{align}
    Z_\mathrm{sL} &= \frac{2\pi}{\sqrt{k}} \frac{2\pi}{b'} \Gamma(-2s') \left(\frac{\mu}{2b'^2}\right)^{2s'} \frac{\Gamma(2s'+1)}{\Gamma(s'+1)^2} \int \mathcal{D}\hat{r}' \mathcal{D}\hat{\theta}' e^{-S_{\mathrm{LD\times}S^1}[\hat{r}',\hat{\theta}']} \nonumber \\
    &\quad \times W_+(0)\, W_+(1)\, W_-(\infty) \left(\int d^2z \, W_+(z,\Bar{z})\right)^{s'-2} \left(\int d^2z \, W_-(z,\Bar{z})\right)^{s'-1}\label{gauge fixed sine-Liouville partition function},
\end{align}
The partition function of the sine-Liouville string theory is given by the product of three partition functions:
\begin{align}
    Z_\mathrm{string}=Z_\mathrm{sL}Z_MZ_c \ .
\end{align}
Here, $Z_M$ is the partition function of the internal CFT\@.

Let us study where in target space the closed string is inserted by the vertex operator $W_\pm$. For this, recall the operator product expansion (OPE) in the $\mathrm{LD\times}S^1$ CFT:
\begin{align}
    \hat{r}(z,\Bar{z}) \hat{r}(w,\Bar{w}) &\sim -\frac{1}{2} \ln|z-w|^2, \\
    \hat{\theta}_L(z) \hat{\theta}_L(w) &\sim -\frac{1}{2} \ln(z-w), \\
    \hat{\theta}_R(\Bar{z}) \hat{\theta}_R(\Bar{w}) &\sim -\frac{1}{2} \ln(\Bar{z}-\Bar{w}).
    \label{operator product expansion}
\end{align}
We can easily verify
\begin{align}
    \lim_{(z,\Bar{z}) \to (w,\Bar{w})} \hat{r}(z,\Bar{z}) W_\pm(w,\Bar{w}) &= \lim_{\rho \to 0} 2b_{sL} \ln\rho \cdot W_\pm(w,\Bar{w}), 
 \label{rhat-W_OPE}
\\
    \lim_{(z,\Bar{z}) \to (w,\Bar{w})} \hat{\theta}(z,\Bar{z}) W_\pm(w,\Bar{w}) &= \pm \sqrt{k} \phi W_\pm(w,\Bar{w}),
\label{thetahat-W_OPE}
\end{align}
where we defined $z-w =: \rho e^{i\phi}$ and $\Bar{z}-\Bar{w} =: \rho e^{-i\phi}$. These show that the asymptotic closed string states prepared by $W_\pm$ are injected from or run off to $\hat{r} \to -\infty$, winding around the thermal circle as $\hat{\theta} \sim \sqrt{k} \phi$. This will be important when we specify the profile of a folded string worldsheet in the next section.

\subsection{Folded Strings and String Entanglement Entropy}
\label{ss:folded strings and string entanglement entropy}

We saw that the sine-Liouville partition function $Z_{\rm sL}$ can be
written as a string amplitude of $2s'$ closed strings.  In this section,
we will see that, by switching to the open string channel, $Z_{\rm sL}$
can be interpreted as the partition function for $2s'$ (folded) open
strings.  These open strings naturally split into two groups of $s'$
strings each and, upon continuation to Lorentzian signature, each group
propagates into a separate copy of flat spacetime. Then it is quite
natural to introduce entanglement entropy between the two groups of
strings and interpret it as the sine-Liouville dual of the thermal
black-hole entropy on the cigar side.  For simplicity, we first discuss
the case of $s'=1$ and then proceed to more general cases with $s'\ge
2$.\footnote{In order to fix the position of three operators, we really
need $s'\ge 3$. However, for the sake of argument, we start with the
case of \( s' = 1 \).}

For \( s' = 1 \), we can assume that the worldsheet positions of \( W_+
\) and \( W_- \) are \( z = 0 \) and \( z = \infty \), respectively, and
parameterize the worldsheet coordinate as \( z = \rho e^{i\phi} \).  If
we interpret the radial coordinate \( \rho \) as time (radial
quantization), the closed string inserted by \( W_+ \) at \( \rho = 0 \)
propagates over time and gets annihilated at \( \rho = \infty \), where
\( W_- \) is inserted.  In the target space picture, a winding closed
string is sent in from \( \hat{r} = -\infty \), turns around at some
finite value of \( \hat{r} \), and then goes back to \( \hat{r} =
-\infty \). This ``closed string channel'' description is illustrated in
Fig.\ \ref{closed string channel}, where the blue-colored sheet
represents the worldsheet.

\begin{figure}[htbp]
\begin{center}
\includegraphics[height=50mm]{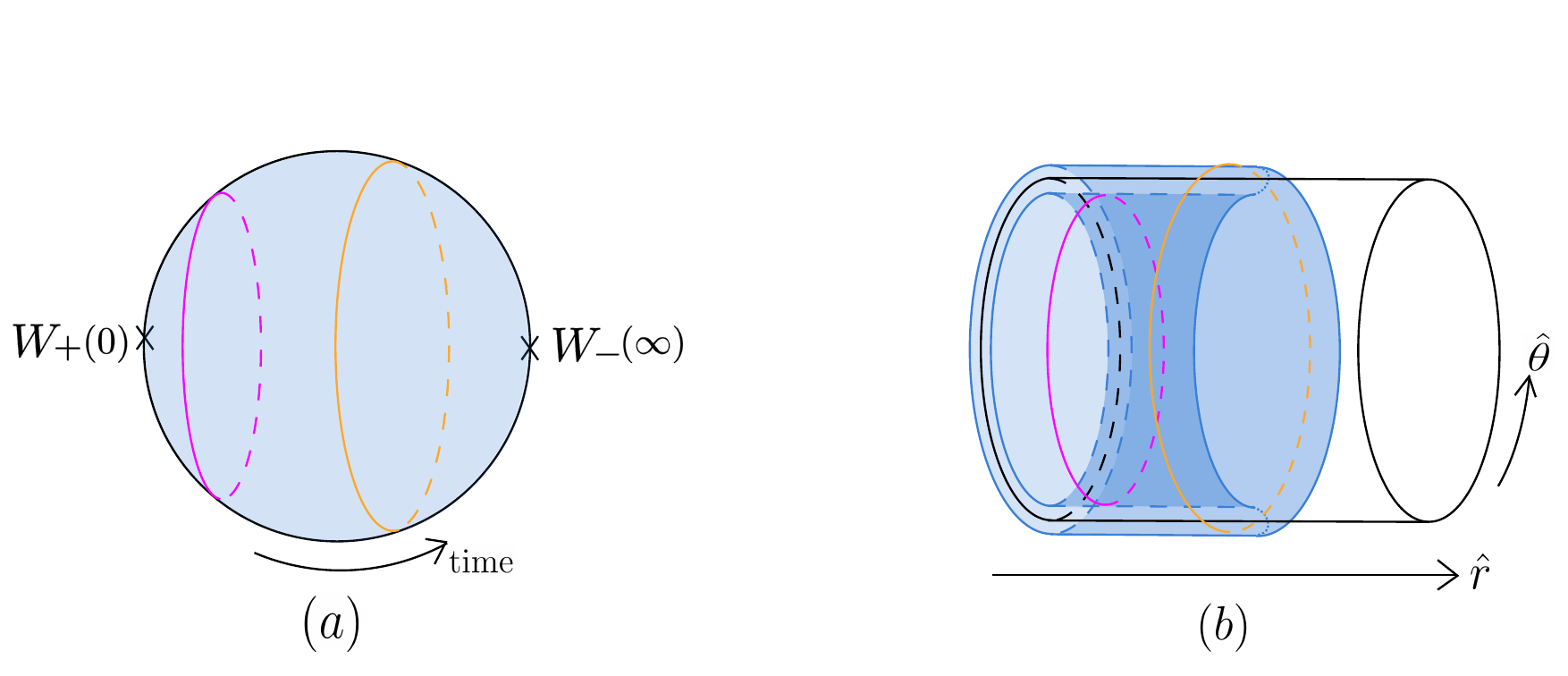}
 \caption{\it The closed string channel for $s'=1$. (a) The worldsheet
 picture. The magenta and orange lines represent the worldsheet on two different time slices
 (constant-$\rho$ curves). (b) The spacetime picture. The winding
 closed string comes in from $\hat{r}=-\infty$, turns around at some
 value of $\hat{r}$, and goes back to $\hat{r}=-\infty$.  The worldsheet at the two time slices are also shown.}  \label{closed
 string channel}
\end{center}
\end{figure}

\begin{figure}[htbp]
\begin{center}
\includegraphics[height=50mm]{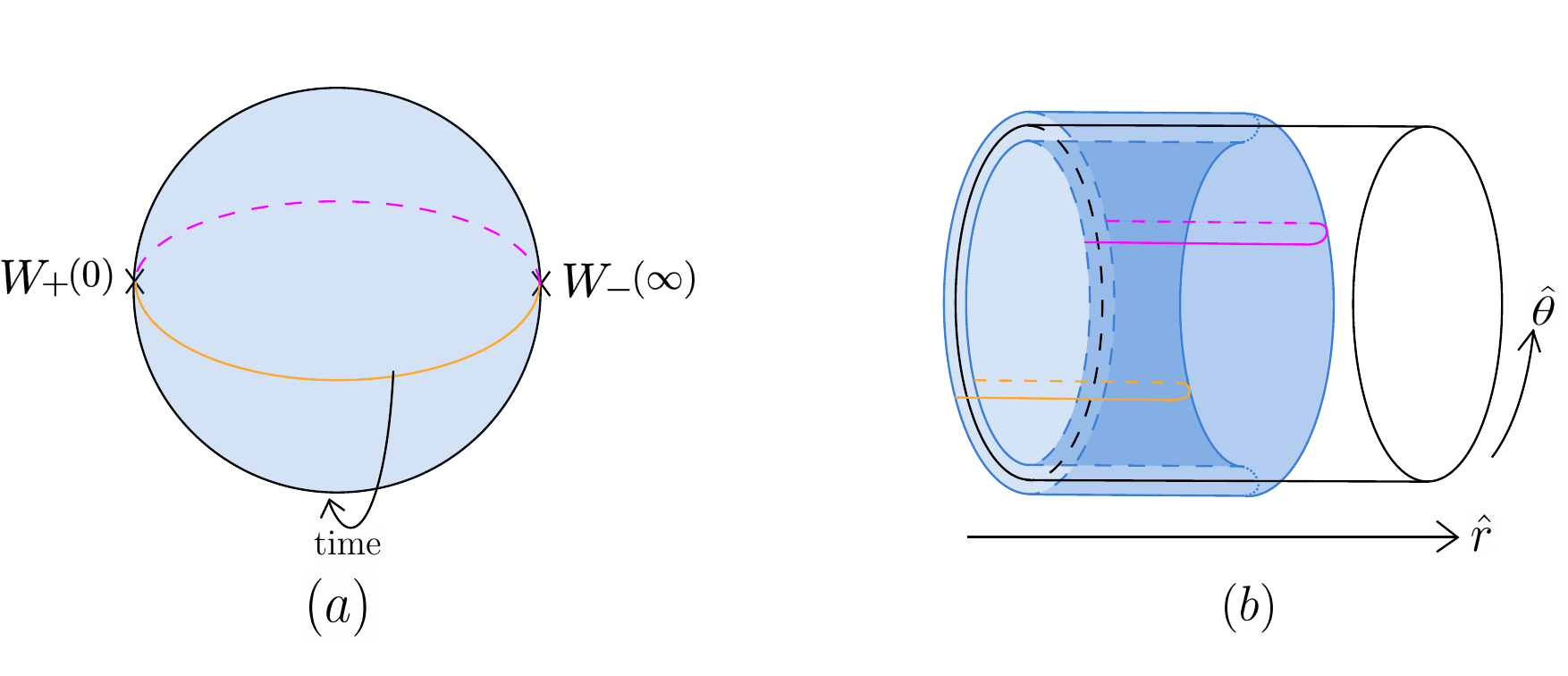}
 \caption{\it The open string channel for $s'=1$. (a) The worldsheet
 picture.  The magenta and orange lines represent the worldsheet at two different time slices
 (constant-$\phi$ curves). (b) The spacetime picture.  A folded open
 string extending along $\hat{r}$ propagates around the $\hat{\theta}$
 circle. The worldsheets corresponding to those in the worldsheet picture are also shown.}  \label{open string channel}
\end{center}
\end{figure}

Alternatively, we may take the angular variable $\phi$ as time (angular
quantization). In this ``open string channel'' description, a time slice
(constant-$\phi$ line) on the worldsheet represents an open string, with
the vertex operators $W_\pm$ giving the boundary condition at its
endpoints~\cite{Agia:2022srj}.  In the target space picture, we
have an open string that extends from $\hat{r}= -\infty$ to some finite
$\hat{r}$ where it folds back, continuing toward $\hat{r}=
-\infty$. Such an open string is called a ``folded string''
\cite{Jafferis:2021ywg}.  
The OPE \eqref{thetahat-W_OPE} says that the folded string propagates once
around the $\hat{\theta}$ circle as we go once around the vertex operator $W_\pm$ on the
worldsheet, and that near the vertex operators we can identify the worldsheet time~$\phi$ with the spacetime Euclidean time $\hat{\theta}$ by $\hat{\theta}=\sqrt{k}\phi$; see Fig.\ \ref{open string channel}.  This identification $\hat{\theta}=\sqrt{k}\phi$ is not correct away from the vertex operators, but for simplicity of argument, we pretend that this is true for $s'=1$; we will make this more precise when we discuss general $s'\ge 2$.

\begin{figure}[htbp]
\begin{center}
\includegraphics[height=60mm]{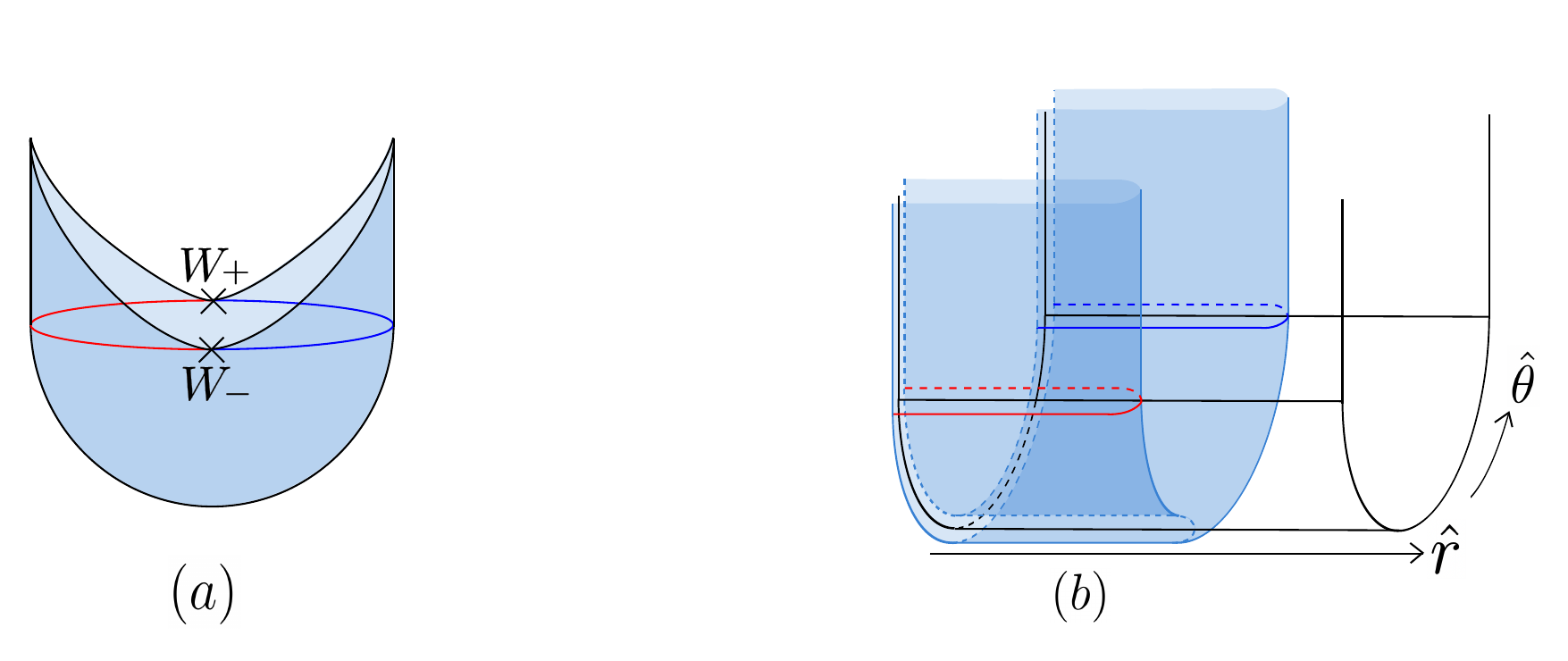}
 \caption{\it The pair of folded open strings to propagate in Lorentzian
 continuation, for $s'=1$. (a) The pair of open strings in the worldsheet
 picture. The Euclidean, hemispherical worldsheet with two vertices inserted prepares a pair of open strings, which propagate into the Lorentzian part of the worldsheet shown as vertical wedges. (b) The target space picture. The Euclidean half-cylinder spacetime is connected to two separate copies of flat Lorentzian space (the vertical parts). The red and blue lines
 represent the folded open strings on $L$ ($\hat{\theta}=0$) and $R$ ($\hat{\theta}=\pi\sqrt{k}$), respectively.  
 \label{folded string s=1}}
\end{center}
\end{figure}

Now, let us consider continuation to Lorentzian time.  On the dual cigar
side, a Euclidean half-cigar prepares the Hartle-Hawking state in the
Lorentzian, two-sided black hole spacetime, and the thermal black-hole
entropy is nothing but the entanglement entropy between the left and
right sides of the Lorentzian black hole.  By the FZZ duality, the
sine-Liouville dual of this Hartle-Hawking state must be obtained
by a Schwinger-Keldysh contour \cite{Jafferis:2021ywg}; namely, we take the
Euclidean half-circle $\hat{\theta}\in[0,\pi\sqrt{k}]$ and continue it
at $\hat{\theta}=0$ and $\hat{\theta}=\pi\sqrt{k}$ to two copies of
Lorentzian spacetime.  If we denote the $\hat{\theta}=0$ and
$\hat{\theta}=\pi\sqrt{k}$ slices by $L$ and $R$, respectively, then the
open string in the Euclidean spacetime is continued to a pair of open
strings propagating in the two disconnected, Lorentzian spacetimes
(Fig.~\ref{folded string s=1}). Because these strings are all there are
on the sine-Liouville side, it is natural to expect that the
entanglement entropy between them is equal to the thermal black-hole entropy
on the cigar side.

In the $s'\ge 2$ case, the essential idea is the same but the situation
is richer and more interesting.  In the closed string channel, there are
$s'$ $W_+$ vertex operators where closed strings are created and $s'$
$W_-$ vertex operators where they are annihilated.  In the target space
picture, $s'$ winding closed strings are injected from
$\hat{r}=-\infty$.  As these strings propagate in spacetime, they all
interact with each other (recall that the worldsheet has the topology of a sphere), turn
around at some point, and then go back to $\hat{r}= -\infty$.

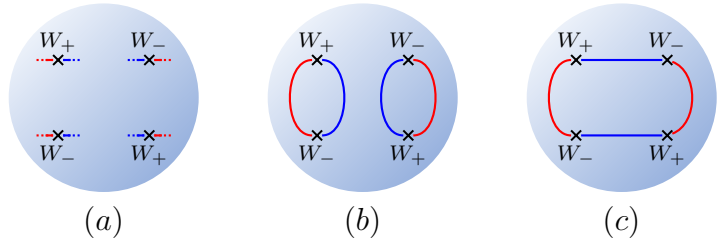
\begin{figure}[tbp]
\begin{center}

\def\tikzball{
  \tikzset{xcross/.style={
      draw,shape=cross out,inner sep=0pt,minimum size=3pt,line width=0.75pt}}
   \shade[shading=axis,               
         left color=ws1!50!white,        
         right color=ws2, shading angle=45]       
    (0,0) circle (2.5cm);
 \node[xcross] (nw) at (-1.2,1) {};
 \node[xcross] (ne) at (1.2,1) {};
 \node[xcross] (sw) at (-1.2,-1) {};
 \node[xcross] (se) at (1.2,-1) {};
 }
 \begin{tabular}{c@{\hspace{5ex}}c@{\hspace{5ex}}c}
\begin{tikzpicture}[scale=0.5]
\tikzball
\node[anchor=south,font=\scriptsize,yshift=-3] at (nw.north) {$W_+$};
\node[anchor=south,font=\scriptsize,yshift=-3] at (ne.north) {$W_-$};
\node[anchor=north,font=\scriptsize,yshift=2] at (sw.south) {$W_-$};
\node[anchor=north,font=\scriptsize,yshift=2] at (se.south) {$W_+$};
\draw[color=blue,thick,densely dotted] (nw.east) -- +(0.5,0);
\draw[color=blue,thick] (nw.east) -- +(0.2,0);
\draw[color=red,thick,densely dotted] (nw.west) -- +(-0.5,0);
\draw[color=red,thick] (nw.west) -- +(-0.2,0);
\draw[color=red,thick,densely dotted] (ne.east) -- +(0.5,0);
\draw[color=red,thick] (ne.east) -- +(0.2,0);
\draw[color=blue,thick,densely dotted] (ne.west) -- +(-0.5,0);
\draw[color=blue,thick] (ne.west) -- +(-0.2,0);
\draw[color=blue,thick,densely dotted] (sw.east) -- +(0.5,0);
\draw[color=blue,thick] (sw.east) -- +(0.2,0);
\draw[color=red,thick,densely dotted] (sw.west) -- +(-0.5,0);
\draw[color=red,thick] (sw.west) -- +(-0.2,0);
\draw[color=red,thick,densely dotted] (se.east) -- +(0.5,0);
\draw[color=red,thick] (se.east) -- +(0.2,0);
\draw[color=blue,thick,densely dotted] (se.west) -- +(-0.5,0);
\draw[color=blue,thick] (se.west) -- +(-0.2,0);
\end{tikzpicture}
&
\begin{tikzpicture}[scale=0.5]
\tikzball
\node[anchor=south,font=\scriptsize,yshift=-3] at (nw.north) {$W_+$};
\node[anchor=south,font=\scriptsize,yshift=-3] at (ne.north) {$W_-$};
\node[anchor=north,font=\scriptsize,yshift=2] at (sw.south) {$W_-$};
\node[anchor=north,font=\scriptsize,yshift=2] at (se.south) {$W_+$};
 \draw[color=red,thick] (nw) to [out=180, in=180] (sw);
 \draw[color=blue,thick] (nw) to [out=0, in=0] (sw);
 \draw[color=blue,thick] (ne) to [out=180, in=180] (se);
 \draw[color=red,thick] (ne) to [out=0, in=0] (se);
\end{tikzpicture}
&
\begin{tikzpicture}[scale=0.5]
\tikzball
\node[anchor=south,font=\scriptsize,yshift=-3] at (nw.north) {$W_+$};
\node[anchor=south,font=\scriptsize,yshift=-3] at (ne.north) {$W_-$};
\node[anchor=north,font=\scriptsize,yshift=2] at (sw.south) {$W_-$};
\node[anchor=north,font=\scriptsize,yshift=2] at (se.south) {$W_+$};
 \draw[color=blue,thick] (nw) to [out=0, in=180] (ne);
 \draw[color=red,thick] (nw) to [out=180, in=180] (sw);
 \draw[color=blue,thick] (sw) to [out=0, in=180] (se);
 \draw[color=red,thick] (ne) to [out=-20, in=20] (se);
\end{tikzpicture}
\\
$(a)$&$(b)$&$(c)$
\end{tabular}
  \caption{\it The location of the $L$ and $R$ curves on a spherical worldsheet
  for $s'=2$. (a): The OPE fixes the curves only near $W_\pm$
  insertions. (b) and (c): two possible ways to connect $L$ and $R$
  curves.  } \label{fig:L and R curves s=2}
\end{center}
\end{figure}

In the open string channel, one might think of taking the worldsheet angle
around each vertex operator as worldsheet time, but that would be valid
only near the vertex operators.  Instead, for our purposes, it is more
convenient to consider worldsheet curves along which $\hat{\theta}=0$ or
$\hat{\theta}=\pi\sqrt{k}$; namely, these curves are the sets of points
where the string passes the $L$ and $R$ slices in spacetime (we must have really done this also for $s'=1$).  The OPE
\eqref{thetahat-W_OPE} tells us where these $L$ and $R$ curves are very
near the insertions (Fig.~\ref{fig:L and R curves s=2}(a)), but their
precise form away from the insertions and how they are connected to each
other depends on the actual functional form of
$\hat{\theta}(z,\bar{z})$.\footnote{Even for $s'=1$, we should really use the $L$ and $R$ curves instead of the constant-$\phi$ curves, because it is the former along which Lorentzian continuation is done.}

Depending on the specific functional form of $\hat{\theta}(z,\bar{z})$,
there are different ways to connect $L$ and $R$ curves; Fig.~\ref{fig:L
and R curves s=2}(b) and (c) show examples of such different
possibilities.\footnote{There are more general possibilities for the $L$
and $R$ curves; for example, we can have closed loops that are not
ending on any vertex operator.  However, we will discuss such more
general possibilities in the next section and here we will focus on
basic examples for the sake of presentation.}  Because a pair of $L$ and
$R$ curves emanates from each of $2s'$ vertex operators, there are $s'$
$L$ curves and $s'$ $R$ curves in total.  On the $L$ and $R$ curves is
an entangled state of $2s'$ open strings and, upon Lorentzian
continuation, the $s'$ strings on $L$ propagate into a copy of flat
spacetime, while the $s'$ strings on $R$ propagate into a distinct copy
of flat spacetime.  Just as in the $s'=1$ case, the entanglement entropy
between the $L$ and $R$ groups of strings is naturally expected to be the
dual of the entropy of the cigar black hole.

\begin{figure}[htbp]
\begin{center}
 \includegraphics[height=120mm]{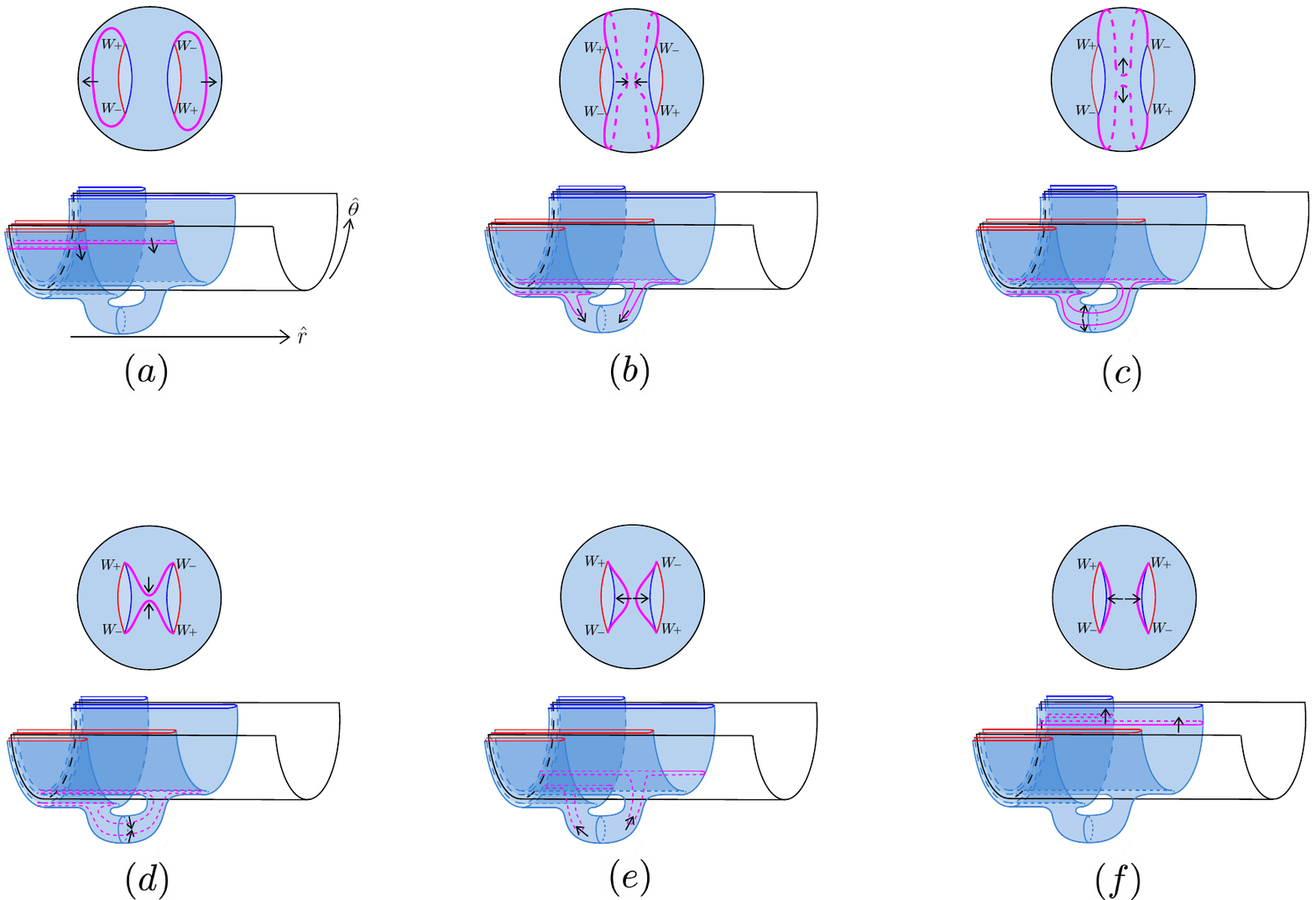}
 \caption{\it The worldsheet and spacetime pictures of string interaction corresponding to the situation in Fig.~\ref{fig:L and R curves s=2}(b).  
Such Euclidean processes contribute to the entangled state of the pairs of open strings that propagate into the Lorentzian spacetime.
 (a) A pair of folded strings starts from the $\hat\theta=0$ line (the red curves).  The string pair is drawn in magenta. (b) Parts of the string pair approach each other.  (c) After reconnection, the string pair moves away from each other. (d) The string pair is about to reconnect again.  (e) After the second reconnection, the string pair moves away from each other.  (f) The strings approach the $\hat{\theta}=\pi\sqrt{k}$ line (the blue curves).
Again, for illustrative simplicity, the spacetime wormhole and the strings in it are drawn as if they went out of the 2d surface of the $\hat{r}$-$\hat{\theta}$ cylinder, but strictly speaking they all lie inside the $\hat{r}$-$\hat{\theta}$ cylinder.  This process can be thought of as the two folded open strings exchanging a closed string.
 \label{folded string s=2}}
\end{center}
\end{figure}

Different ways to connect $L$ and $R$ curves give different ways to
prepare entangled string states, with different ways of interaction.
As illustrative examples, in Figs.~\ref{muiltiple folded strings is
sine-Liouville} and \ref{folded string s=2}, we give the worldsheet and spacetime picture of the string time
evolution and interaction corresponding to the situation in Fig.~\ref{fig:L and R curves
s=2}(b).

\begin{figure}[htbp]
\begin{center}
\includegraphics[height=60mm]{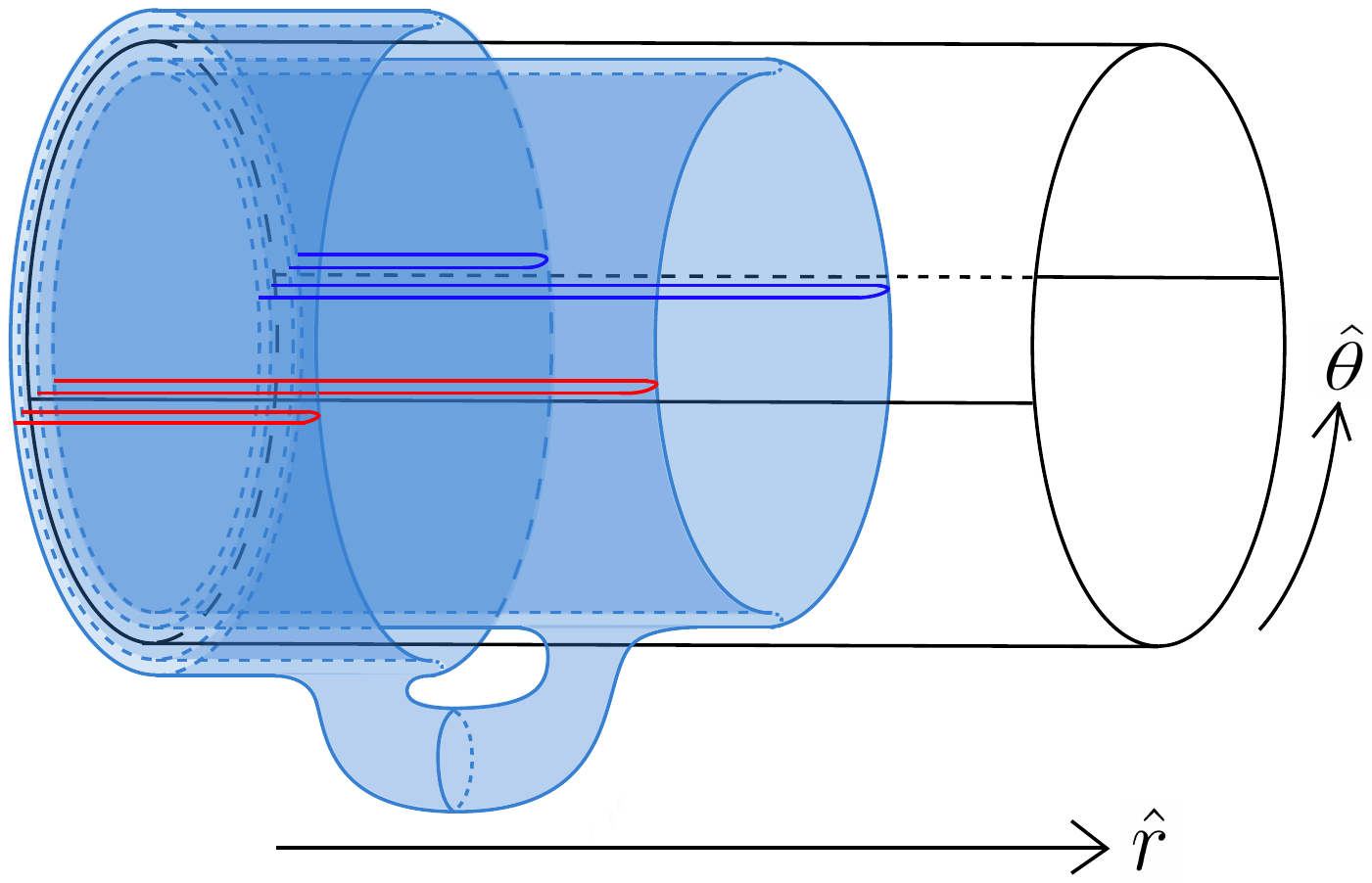}
 \caption{\it The spacetime picture of the string time evolution shown 
 in Fig.~\ref{fig:L and R curves s=2}(b).  We can think of the spherical worldsheet in Fig.~\ref{fig:L and R curves s=2}(b) as two hemispheres---left and right---glued together along a great circle of longitude.  Topologically, this is the same as two spherical worldsheets connected by a throat, whose spacetime picture is two copies of the worldsheet shown in Fig.~\ref{open string channel} connected by a wormhole as shown here.  
For illustrative simplicity, the spacetime wormhole is drawn as if it went out of the 2d surface of the $\hat{r}$-$\hat{\theta}$ cylinder, but strictly speaking it lies inside the $\hat{r}$-$\hat{\theta}$ cylinder.
 }  \label{muiltiple folded strings
 is sine-Liouville}
\end{center}
\end{figure}

\section{String Entanglement in Sine-Liouville String}
\label{sec:string_entanglement_in_sL_string}

\subsection{Review of \cite{Halder:2024gwe}}
\label{sec:review_of_Halder-Jafferis}

Here we review the approach of \cite{Halder:2024gwe} to evaluating the thermal entropy of the cigar black hole, in order to contrast their approach with ours.  

The black hole entropy can be computed from the Euclidean spacetime partition function~\cite{Gibbons:1976ue}, which is given in principle by the exponential of the string partition function.  By the FZZ duality, the string partition function for the cigar black hole background is equal to the string partition function for the sine-Liouville background, but Ref.~\cite{Halder:2024gwe} showed that the latter vanishes on $S^2$.  To circumvent this problem, Ref.~\cite{Halder:2024gwe} instead used the relation between thermal entropy and the partition function in a conical deficit geometry~\cite{Susskind:1994sm, Lewkowycz:2013nqa}.  Namely, one introduces a deficit angle $\delta$ at the tip of the cigar in the Euclidean black-hole background, and relates the $\delta$ derivative of its spacetime partition function to thermal entropy~\cite{Halder:2024gwe}.  Again, the relevant spacetime partition function is the exponential of the worldsheet partition function, which can be evaluated using the sine-Liouville string (and the thermal entropy computed from it turns out to be non-vanishing).

The spacetime partition function with a conical singularity at the tip of the cigar with deficit angle $\delta$ can be related to the $n$-th R\'enyi entropy with $n=1-\delta$ as follows:
\begin{align}
    \mathcal{S}_n^{\rm th} = \frac{1}{1-n} \ln \frac{\mathcal{Z}_\mathrm{cigar}(n)}{\left(\mathcal{Z}_\mathrm{cigar}(1)\right)^n}, \label{cigar_nth_renyi_entropy}
\end{align}
where $\mathcal{Z}_\mathrm{cigar}(1)$ is the spacetime partition function of the cigar geometry without conical deficit, for which the periodicity of the $\theta$ direction is $2\pi$, while $\mathcal{Z}_\mathrm{cigar}(n)$ is the spacetime partition function with the $\theta$-periodicity changed to $2\pi n$.
Because $\mathcal{Z}_\mathrm{cigar} = \exp(Z_\mathrm{cigar})$ where $Z_\mathrm{cigar}$ is  the worldsheet partition function, the thermal R\'enyi entropy~\eqref{cigar_nth_renyi_entropy} can be expressed as
\begin{align}
    \mathcal{S}_n^{\rm th} = \frac{1}{1-n} \left(Z_\mathrm{cigar}(n) - n Z_\mathrm{cigar}(1)\right).
    \label{cigar_nth_renyi_entropy_ito_roman_Z}
\end{align}
By the FZZ duality, we have $Z_\mathrm{cigar}=Z_\mathrm{sL}$ and therefore 
\begin{align}
    \mathcal{S}_n^{\rm th} =\frac{1}{1-n} \left(Z_\mathrm{sL}(n) - n Z_\mathrm{sL}(1)\right),
    \label{sine_Liouville_nth_renyi_entropy}
\end{align}
where $Z_{\rm sL}(n)$ is the sine-Liouville string partition function with $\hat{\theta}$-periodicity $2\pi n\sqrt{k}$.

The thermal entropy is then obtained by taking the $n \to 1$ ($\delta\to 0$) limit of the R\'enyi entropy.  Instead of changing the periodicity of the $\hat\theta$ direction, Ref.~\cite{Halder:2024gwe} introduced a conical singularity in the sine-Liouville description by modifying the vertex operator $W_\pm$, at leading order in $\delta$,  as follows:
\begin{align}
    W_\pm = e^{-2b_\mathrm{sL} \hat{r}} e^{\pm i\sqrt{k}(\hat{\theta}_\mathrm{L} - \hat{\theta}_\mathrm{R})} 
    \xrightarrow{~\sqrt{k}\to (1-\delta)\sqrt{k}~}
    e^{-2b_\mathrm{sL} \left(1-\frac{\delta}{1-\frac{2}{k}}\right)\hat{r}} 
    e^{\pm i\sqrt{k}(1-\delta)(\hat{\theta}_\mathrm{L} - \hat{\theta}_\mathrm{R})},
\end{align}
while keeping the periodicity of the $\hat\theta$ direction unchanged. This  effectively changes the circumference of the $S^1_{\hat\theta}$ direction as $2\pi\sqrt{k}\to 2\pi\sqrt{k}(1-\delta)$.  
Evaluating $Z_{\rm sL}$ in \eqref{gauge fixed sine-Liouville partition function}
with this modified vertex operator and using  the relation
\eqref{sine_Liouville_nth_renyi_entropy}, Ref.~\cite{Halder:2024gwe} derived the thermal entropy of the cigar black hole and claimed that it reproduces the known $\alpha'$ corrections computed using the relation \cite{Chen:2021emg} between the cigar black hole and the large-$D$ Schwarzschild black hole \cite{Callan:1988hs}.

We emphasize that the approach of \cite{Halder:2024gwe} is to evaluate the \emph{spacetime thermal} entropy of the cigar black hole using the sine-Liouville string.  This is in contrast with our approach, which is to relate the cigar black hole entropy to the \emph{string entanglement} entropy of the sine-Liouville string.

\subsection{String Entanglement Entropy in Sine-Liouville}
\label{ss:ws_ee_in_sL}

In section \ref{ss:folded strings and string entanglement entropy}, we argued that the thermal entropy of the cigar black hole is equal, on the sine-Liouville side, to the string entanglement entropy between two groups of $s'$ open strings each, which are on constant-$\hat{\theta}$ curves (that we call $L$ and $R$ curves) and propagate into two disconnected copies of flat spacetime upon Lorentzian continuation. Here we discuss how to evaluate their entanglement entropy, making more precise the rough argument given in section~\ref{ss:folded strings and string entanglement entropy} along the way.

As we discussed in section \ref{ss:folded strings and string entanglement entropy}, on the worldsheet are $2s'$ $W_\pm$ vertex operators connected by $s'$ $L$-curves and $s'$ $R$-curves.
How the vertex operators are connected by those curves depends on the actual functional form of $\hat{\theta}(z,\bar{z})$, as illustrated in Fig.~\ref{fig:L and R curves s=2}(b) and~(c).  These curves do not intersect each other, and on each vertex operator one $L$-curve and one $R$-curve end.
On the union of the $s'$ $L$-curves, we have the Hilbert space $\mathcal{H}_L$ of $s'$ open strings, and on the union of the $s'$ $R$-curves, we have the Hilbert space $\mathcal{H}_R$ of $s'$ open strings.  
The worldsheet path integral defines a pure state $\ket{\Psi}\in \mathcal{H}_L\otimes \mathcal{H}_R$ and the density matrix $\rho_{LR}=\ket{\Psi}\!\bra{\Psi}$. 

\begin{figure}[htbp]
\begin{center}

\begin{align}
\tr_L \rho_L^n=
\left.
\vcenter{\hbox{
\begin{tikzpicture}]
  \tikzset{
    xcross/.style={
      draw,
      shape=cross out,
      inner sep=0pt,
      minimum size=4pt,
      line width=0.5pt,
    }
  }
\draw[fill=ws1] (0,0) -- +(4,0) -- +(5,1) -- +(1,1) -- cycle;
 \node[xcross,label={[font=\scriptsize,yshift=-3]above:$W_+$}] (v1ll) at (1,0.5) {};
 \node[xcross,label={[font=\scriptsize,yshift=-3]above:$W_-$}] (v1lr) at (2,0.5) {};
 \node[xcross,label={[font=\scriptsize,yshift=-3]above:$W_+$}] (v1rl) at (3,0.5) {};
 \node[xcross,label={[font=\scriptsize,yshift=-3]above:$W_-$}] (v1rr) at (4,0.5) {};
 \draw (v1ll) -- (v1lr);
 \draw (v1rl) -- (v1rr);
\draw[fill=ws1] (0,-1.5) -- +(4,0) -- +(5,1) -- +(1,1) -- cycle;
 \node[xcross,label={[font=\scriptsize,xshift=5,yshift=2]below left:$W_+$}] (v2ll) at (1,-1.0) {};
 \node[xcross,label={[font=\scriptsize,xshift=5,yshift=2]below left:$W_-$}] (v2lr) at (2,-1.0) {};
 \node[xcross,label={[font=\scriptsize,xshift=5,yshift=2]below left:$W_+$}] (v2rl) at (3,-1.0) {};
 \node[xcross,label={[font=\scriptsize,xshift=5,yshift=2]below left:$W_-$}] (v2rr) at (4,-1.0) {};
 \draw (v2ll) -- (v2lr);
 \draw (v2rl) -- (v2rr);
\draw[fill=ws1] (0,-3) -- +(4,0) -- +(5,1) -- +(1,1) -- cycle;
 \node[xcross,label={[font=\scriptsize,yshift=2]below:$W_+$}] (v3ll) at (1,-2.5) {};
 \node[xcross,label={[font=\scriptsize,yshift=2]below:$W_-$}] (v3lr) at (2,-2.5) {};
 \node[xcross,label={[font=\scriptsize,yshift=2]below:$W_+$}] (v3rl) at (3,-2.5) {};
 \node[xcross,label={[font=\scriptsize,yshift=2]below:$W_-$}] (v3rr) at (4,-2.5) {};
 \draw (v3ll) -- (v3lr);
 \draw (v3rl) -- (v3rr);
\draw (v1ll) -- (v3ll);
\draw (v1lr) -- (v3lr);
\draw (v1rl) -- (v3rl);
\draw (v1rr) -- (v3rr);
\end{tikzpicture}
}} \right\}n
\end{align}
    \caption{\it Replica trick on the worldsheet.  (As we discuss around \eqref{def_Wn}, the vertex operators $W_\pm$ inserted at the branch points must be appropriately modified when we go to the covering space, although it is not explicitly shown here.)
        \label{fig:stringy_replica}}
\end{center}
\end{figure}

In order to compute the entanglement entropy between the two groups of strings, we introduce the reduced density matrix $\rho_L=\Tr_R [\rho_{LR}]$ where $\Tr_R$ is the trace on $\mathcal{H}_R$ defined on the $R$-curves.  Then, as is standard, $\Tr_L [\rho_L^n]$ can be computed by the replica trick by going to the $n$-sheeted cover of the worldsheet with the $L$ curves being the branch cuts (Fig.~\ref{fig:stringy_replica}).
The quantity 
\begin{align}
  S_n =\frac{1}{1-n} \log {\Tr_L[\rho_L^n]\over (\Tr_L[\rho_L])^n}
    \label{string_Renyi_single-particle}
\end{align}
gives the R\'enyi entropy between the entangled pair of groups of strings. However, this is just the entropy between a single pair of string groups; there can be multiple pairs of string groups.  In other words, \eqref{string_Renyi_single-particle} is the contribution from a connected string diagram but there can be an arbitrary number of such connected diagrams, mutually disconnected, representing multiple entangled pairs of string groups.   Just as going from \eqref{cigar_nth_renyi_entropy} to \eqref{cigar_nth_renyi_entropy_ito_roman_Z}, the ``multi-pair'' R\'enyi entropy that takes this into account is given by 
\begin{align}
\mathcal S_n = \frac{1}{1-n} \left( \Tr_L [\rho_L^n] - n  \Tr_L[ \rho_L] \right).
    \label{string_Renyi_multi-particle}
\end{align}
The string entanglement entropy that we are after is obtained by taking the limit $n \to 1$ of this quantity.

\subsubsection*{Features of the worldsheet replica trick}

Depending on the functional form of $\hat\theta(z,\bar{z})$, the structure of the $L$ and $R$ curves is different (see Fig.~\ref{fig:L
and R curves s=2}(b) and (c) for examples in the $s'=2$ case).  So, it might appear that we must do separate replica computations for different ``sectors'' in the space of functions $\{\hat\theta(z,\bar{z})\}$. However, this is not really so -- any structure of the $L$ and $R$ curves leads to the same replica.   Let us assume that we start with $\rho_{LR}$, take the trace $\rho_L=\Tr_R [\rho_{LR}]$ on $\mathcal{H}_R$ over the $R$ curves, and then go to a covering space with the $L$ curves being the branch cuts.
First, because we integrate over all possible field configurations on the $R$ curves, the location of the $R$ curves is irrelevant and we can continuously deform the $R$ curves without changing the replica computation.   Also, because continuously deforming branch cuts does not change the covering space, we can continuously deform the $L$ curves.

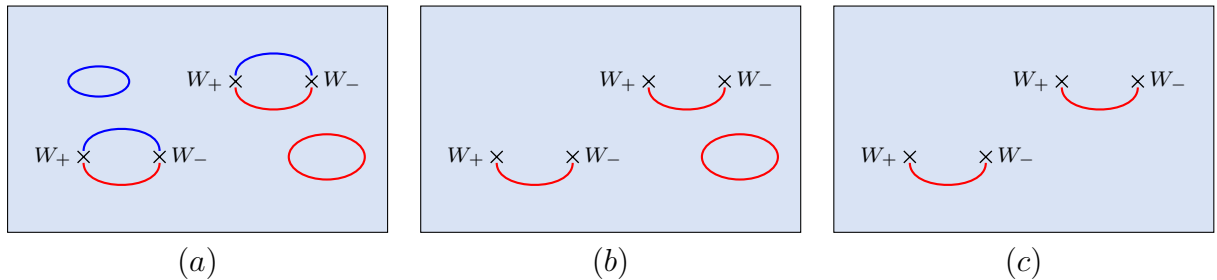
\begin{figure}[tbp]
\begin{center}
\begin{tabular}{ccc}
\begin{tikzpicture}
  \tikzset{xcross/.style={draw,shape=cross out,inner sep=0pt,
      minimum size=4pt,line width=0.5pt}}
\draw[fill=ws1] (0,0) rectangle (5,3);
\node[xcross,label={[font=\scriptsize,xshift=2]left:$W_+$}] (ll) at (1,1) {};
\node[xcross,label={[font=\scriptsize,xshift=-2]right:$W_-$}] (lr) at (2,1) {};
\node[xcross,label={[font=\scriptsize,xshift=2]left:$W_+$}] (rl) at (3,2) {};
\node[xcross,label={[font=\scriptsize,xshift=-2]right:$W_-$}] (rr) at (4,2) {};
\draw[color=blue,thick] (ll) to [out=90, in=90] (lr);
\draw[color=red,thick] (ll) to [out=-90, in=-90] (lr);
\draw[color=blue,thick] (rl) to [out=90, in=90] (rr);
\draw[color=red,thick] (rl) to [out=-90, in=-90] (rr);
\draw[color=blue,thick] (1.2,2) ellipse (4mm and 2mm);
\draw[color=red,thick] (4.2,1) ellipse (5mm and 3mm);
\end{tikzpicture}
&
\begin{tikzpicture}
  \tikzset{xcross/.style={draw,shape=cross out,inner sep=0pt,
      minimum size=4pt,line width=0.5pt}}
\draw[fill=ws1] (0,0) rectangle (5,3);
\node[xcross,label={[font=\scriptsize,xshift=2]left:$W_+$}] (ll) at (1,1) {};
\node[xcross,label={[font=\scriptsize,xshift=-2]right:$W_-$}] (lr) at (2,1) {};
\node[xcross,label={[font=\scriptsize,xshift=2]left:$W_+$}] (rl) at (3,2) {};
\node[xcross,label={[font=\scriptsize,xshift=-2]right:$W_-$}] (rr) at (4,2) {};
\draw[color=red,thick] (ll) to [out=-90, in=-90] (lr);
\draw[color=red,thick] (rl) to [out=-90, in=-90] (rr);
\draw[color=red,thick] (4.2,1) ellipse (5mm and 3mm);
\end{tikzpicture}
&
\begin{tikzpicture}
  \tikzset{xcross/.style={draw,shape=cross out,inner sep=0pt,
      minimum size=4pt,line width=0.5pt}}
\draw[fill=ws1] (0,0) rectangle (5,3);
\node[xcross,label={[font=\scriptsize,xshift=2]left:$W_+$}] (ll) at (1,1) {};
\node[xcross,label={[font=\scriptsize,xshift=-2]right:$W_-$}] (lr) at (2,1) {};
\node[xcross,label={[font=\scriptsize,xshift=2]left:$W_+$}] (rl) at (3,2) {};
\node[xcross,label={[font=\scriptsize,xshift=-2]right:$W_-$}] (rr) at (4,2) {};
\draw[color=red,thick] (ll) to [out=-90, in=-90] (lr);
\draw[color=red,thick] (rl) to [out=-90, in=-90] (rr);
\end{tikzpicture}
\\
$(a)$&$(b)$&$(c)$
\end{tabular}
\end{center}
\caption{\it Loops of $L$ and $R$ curves do not matter. (a) L curves (in red) and R curves (in blue) on the worldsheet, containing loops. (b) Upon tracing out states in $\mathcal{H}_R$, we no longer have R curves, including loops.  We are left with L curves with loops.  (c) Because $L$ loops do not change the structure of the covering space, we can forget about them, left only with L curves without loops. 
\label{fig:loope_on_worldsheet}}
\end{figure}

Besides curves that end on vertex operators, there can be L and R curves that form a closed loop.  In spacetime, they correspond to closed strings propagating into $L$ or $R$ spacetime. Closed $R$ curves on the worldsheet are irrelevant because they disappear when we trace out states in $\mathcal{H}_R$.  On the other hand, when we go into a closed $L$ loop, we go from sheet $i$ to sheet $i+1$ (or $i-1$), but we come back to sheet $i$ when we go back out.  So, such a closed loop only means to ``rename'' some part of sheet $i+1$ to sheet $i$, but creates no non-trivial cycles.  Therefore, such closed $L$ loops does not affect the structure of the covering space.
See Fig.~\ref{fig:loope_on_worldsheet} for illustration.

This means that we can freely deform and reconnect $L$ and $R$ curves.\footnote{Note that the curves have orientation, determined by the direction along which the $\hat{\theta}$ increases.  Reconnection must be consistent with the orientation.}  Using this, one can show that there is no monodromy when one vertex operator goes around another vertex operator past a branch cut.  This is necessary for the integration over the positions of vertex operators to make sense.

So, the replica computation does not depend on the structure of the $L$ and $R$ curves; it only depends on the positions of the $W_\pm$ operators and not on how they are connected by the curves. However, the spacetime shape of the string worldsheet does depend on the curves, as illustrated in
Fig.~\ref{muiltiple folded strings is sine-Liouville}.  In other words, a single covering space captures all possible
states of strings that propagate into $L$ and $R$ spacetimes, including all Euclidean processes that prepare the states, and including an arbitrary number of closed strings.  We find this quite striking.

One other thing to note is that the worldsheet R\'enyi and entanglement entropies are free from UV divergences that come from the region near branch points.  Such UV divergences are present for general CFTs and proportional to the central charge $c$ of the CFT, but for the worldsheet CFT the total central charge vanishes, $c=0$, and the UV divergences are absent.

\subsubsection*{Replica with vertex insertions}

The replica computation we are doing is a non-standard one in that we have vertex operators $W_\pm$ inserted at the branch points. In the closed string channel, they specify what closed string states are inserted.  In the open string channel, their role is to provide boundary conditions for open string endpoints, and that boundary condition defines the open string Hilbert space and the open string Hamiltonian with which to evolve in the angular direction around the insertion \cite{Agia:2022srj}.\footnote{This angular direction is defined only near the insertion, but that is enough for the present discussion.} In the original space, the angular direction is $2\pi$ periodic and we evolve with the angular Hamiltonian for angular time $2\pi$. Going to the $n$-cover to compute $\Tr_L[\rho_L^n]$ means to use the same original boundary condition on each sheet and evolve with the same angular Hamiltonian, but for angular time $2\pi n$.  As we explain in Appendix \ref{app:boundary conditions}, this means that we need to replace the vertex operator as
\begin{align}
    W_\pm = e^{-2b_{\rm sL}\hat{r}}
    e^{\pm i\sqrt{k}(\hat{\theta}_L-\hat{\theta}_R)}
    ~~\to~~
    W_\pm^{(n)} := e^{-2n b_{\rm sL}\hat{r}}
    e^{\pm in\sqrt{k}(\hat{\theta}_L-\hat{\theta}_R)}
    \label{def_Wn}
\end{align}
on the $n$-cover.

The covering space is an $n$-sheeted cover of the complex plane branched at the $2s'$ points where the vertex operators are inserted. This is a Riemann surface of genus $g=(n-1)(s'-1)$. In the linear dilaton CFT, the neutrality condition dictates the number of $W_\pm$ insertions, depending on the genus of the worldsheet.  As we discuss in Appendix \ref{app:neutrality condition}, the change \eqref{def_Wn} is precisely what is needed for the neutrality condition to continue to be satisfied on the $n$-cover for arbitrary $n$.

If we start with the base space with metric $d^2s = dz d\bar{z}$ and go to the covering space branched at $2s'$ insertions, there will be conical singularities at the insertions due to the periodicity of the angle around them being $2\pi n$.  Locally, these singularities can be smoothed out by the coordinate transformation $w = z^{1/n}$ and a Weyl rescaling. This Weyl rescaling does not change the partition function because the total central charge of the worldsheet CFT is zero, $c = 0$.  Thus, the result is a smooth Riemann surface of genus $g = (n-1)(s'-1)$ with $2s'$ $W_\pm^{(n)}$ insertions. The $2s'$-point function on this surface gives $\Tr_L[\rho_L^n]$ in principle.

\subsubsection*{Structure of the string entanglement entropy}

We are interested in computing the entanglement entropy, which is the $\delta=1-n\to 0$ limit of the R\'enyi entropy \eqref{string_Renyi_multi-particle}.  At $\cO(\delta)$, there are two contributions to $\Tr_L [\rho_L^n]$:
\begin{align}  
 \Tr_L[\rho_L^{1-\delta}]
 =\Tr_L[\rho_L]
 +
 \sum_{i=1}^{2s'} 
 \left(
 \begin{minipage}{20ex}contribution from change in $W_\pm(z_i,\bar{z}_i)$\end{minipage}\right)
 +
 \left(\begin{minipage}{15ex}contribution from change in the world\-sheet\end{minipage}\right)
 +\cO(\delta^2).
   \label{order_delta_contrib_to_tr_rho_n}
\end{align}  
The second and third terms are of $\cO(\delta)$.
The second term is coming from the change in the vertex operators
\begin{align}
    W_\pm ~~\to~~W_\pm ^{(1-\delta)} =:  W_\pm ^\delta
        =W_\pm + \delta W_\pm        ,
\end{align}
without going to the covering space.  Let us call this the \emph{vertex operator contribution}.  The third term in \eqref{order_delta_contrib_to_tr_rho_n} is the contribution by going to the covering space of genus $g=-(s'-1)\delta$ keeping the vertex operators unchanged. Let us call this the \emph{replica contribution}.  
Schematically, \eqref{order_delta_contrib_to_tr_rho_n} can be 
depicted as follows:
\begin{equation}
 \Tr_L[\rho_L^{1-\delta}]
 =\underbrace{\,\Tr_L[\rho_L]
+ ~~~\overbrace{\!\!\!\!\!\!
\vcenter{\hbox{\begin{tikzpicture}[scale=1.2]
\draw[fill=ws1] (0,0) -- (2,0) -- (2.5,1) -- (0.5,1) -- cycle;
\draw (0.5,0.45) -- +(0.1,0.1);
\draw (0.5,0.55) -- +(0.1,-0.1) node [above] {\tiny $\delta W_+$};
\draw (1.0,0.45) -- +(0.1,0.1);
\draw (1.0,0.55) -- +(0.1,-0.1) node [above] {\tiny $W_-$};
\draw (1.5,0.45) -- +(0.1,0.1);
\draw (1.5,0.55) -- +(0.1,-0.1) node [above] {\tiny $W_+$};
\draw (2.0,0.45) -- +(0.1,0.1);
\draw (2.0,0.55) -- +(0.1,-0.1) node [above] {\tiny $W_-$};
\end{tikzpicture}}}
\!\!+\cdots+\!\!
\vcenter{\hbox{\begin{tikzpicture}[scale=1.2]
\draw[fill=ws1] (0,0) -- (2,0) -- (2.5,1) -- (0.5,1) -- cycle;
\draw (0.5,0.45) -- +(0.1,0.1);
\draw (0.5,0.55) -- +(0.1,-0.1) node [above] {\tiny $W_+$};
\draw (1.0,0.45) -- +(0.1,0.1);
\draw (1.0,0.55) -- +(0.1,-0.1) node [above] {\tiny $W_-$};
\draw (1.5,0.45) -- +(0.1,0.1);
\draw (1.5,0.55) -- +(0.1,-0.1) node [above] {\tiny $W_+$};
\draw (2.0,0.45) -- +(0.1,0.1);
\draw (2.0,0.55) -- +(0.1,-0.1) node [above] {\tiny $\delta W_-$};
\end{tikzpicture}}}~
}^{\text{vertex operator contribution}}\!\!\!\!\!\!\!}_{\equiv \, Z_\delta}~~
+~~
\overbrace{\!\!\!\!\!
\vcenter{\hbox{\begin{tikzpicture}[scale=1.2]
\draw[fill=ws1] (0,0) -- (2,0) -- (2.5,1) -- (0.5,1) -- cycle;
\draw (0.5,0.45) -- +(0.1,0.1);
\draw (0.5,0.55) -- +(0.1,-0.1) node [above] {\tiny $W_+$};
\draw (1.0,0.45) -- +(0.1,0.1);
\draw (1.0,0.55) -- +(0.1,-0.1) node [above] {\tiny $W_-$};
\draw (1.5,0.45) -- +(0.1,0.1);
\draw (1.5,0.55) -- +(0.1,-0.1) node [above] {\tiny $W_+$};
\draw (2.0,0.45) -- +(0.1,0.1);
\draw (2.0,0.55) -- +(0.1,-0.1) node [above] {\tiny $W_-$};
\draw (0.55,0.5) -- (1.05,0.5);
\draw (1.55,0.5) -- (2.05,0.5);
\node () at (1.95,0.13) {\tiny $\delta$};
\end{tikzpicture}}}~
}^{\text{replica contribution}}\!\!
+\,\cO(\delta^2).
\label{Tr_Zdelta}
\end{equation}
Note that $Z_{\delta}$ contains the contributions from the first and the second terms in the right hand side in \eqref{Tr_Zdelta}. The replica contribution is what is relevant in the standard computation of entanglement entropy using the replica trick, while the vertex operator contribution is new and special to the current situation.

Evaluating the replica contribution is technically challenging, because it requires information about the $2s'$ point function on a Riemann surface of arbitrary genus.  It seems that we also have to know the detail of the internal manifold.  So, in the current paper, we restrict ourselves to evaluating the vertex operator contribution to the string entanglement entropy.  We will come back to the replica contribution and the internal part in the discussion section.

\subsection{Evaluation of String Entanglement Entropy}
\label{Evaluation of the String Entanglement Entropy}

The vertex operator contribution to the string R\'enyi entropy is given by the $2s'$-point function of the integrated $W_{\pm}^\delta$ operators in the $\mathrm{LD}\times S^1$ CFT, times the partition function of the internal CFT,~$Z_M$:
\begin{align}
    Z_\delta &= \frac{1}{\mathrm{vol}(SL(2,\mathbb{C}))}\,\frac{2\pi}{\sqrt{k}} Z_M \frac{2\pi}{b'} \Gamma(-2s') \left(\frac{\mu}{2b'^2}\right)^{2s'} \frac{\Gamma(2s'+1)}{\Gamma(s'+1)^2} 
    \nonumber\\
    &\quad\times\int \mathcal{D}\hat{r}' \mathcal{D}\hat{\theta}' e^{-S_{\mathrm{LD} \times S^1}[\hat{r}',\hat{\theta}']} \prod_{I=1}^{s'}\int d^2z_I \, W^\delta_+(z_I) \prod_{I'=1}^{s'}\int d^2z'_{I'} \, W^\delta_-(z'_{I'})\ .\label{z_delta full}
\end{align}
Here, the prefactor $1/\mathrm{vol}(SL(2,\mathbb{C}))$ is due to the path integral measure of the Polyakov string on a sphere.
In the presence of $W^\delta_\pm$ with a conformal weight
\begin{align}
    h_\delta = b_{sL}(1-\delta)(Q-b_{sL}(1-\delta))+\frac{k}{4}(1-\delta)^2 
    = 1 + (Qb_{sL}-2) \delta
    =: 1 + \epsilon' \ ,
\end{align}
however, the Faddeev-Popov procedure for canceling the divergence from the integral of the location of the marginal operators with $1/\mathrm{vol}(SL(2,\mathbb{C}))$ is inapplicable. See Appendix~\ref{app:String theory with non-marginal vertex operators} for a review of the Faddeev-Popov procedure.
We note that it suffices to evaluate $Z_\delta$ up to linear terms in $\delta$ expansions for the purpose of obtaining the string entanglement entropy. As discussed in Appendix \ref{non-marginal}, the computation of $Z_\delta$ to this order amounts to inserting only a single non-marginal operator $W_+^{2s'\delta}$ with the rest given by the marginal operator $W_{\pm}$.  
Using the prescription in Appendix \ref{app:String theory with non-marginal vertex operators} for making the $2s'$-point function with a non-marginal operator well-defined, we find   
\begin{align}
    Z_{\delta}&= \frac{2\pi}{\sqrt{k}} Z_M Z_c \frac{2\pi}{b'} \Gamma(-2s') \left(\frac{\mu}{2b'^2}\right)^{2s'} \frac{\Gamma(2s'+1)}{\Gamma(s'+1)^2}
    \int \mathcal{D}\hat{r}' \mathcal{D}\hat{\theta}' e^{-S_{\mathrm{LD} \times S^1}[\hat{r}',\hat{\theta}']} W_+(0)\, W_+(1)\, W_-(\infty) \notag\\
    &\quad \times 
     \int d^2z_3 \,|z_3|^{2\epsilon'}\,W^{ 2s'\delta}_+(z_3) \prod_{j=4}^{s'} \int d^2z_j \, W_+ \prod_{k=2}^{s'} \int d^2z_k \, W_- 
     +\mathcal{O}(\delta^2)\ . \label{z_delta_3}
\end{align}
Utilizing the OPEs 
\begin{align}
    &W^\delta_\pm(z,\bar{z})W_{\pm}(w,\bar{w}) \sim |z-w|^{2(1-\delta)}:W_\pm W_{\pm}(w,\bar{w}):\ ,\nonumber\\
    &W^\delta_{\pm}(z,\bar{z})W_{\mp}(w,\bar{w}) \sim |z-w|^{2(1-k)(1-\delta)}:W_{\pm}W_{\mp}(w,\bar{w}):
    \ ,\label{ope-delta}
\end{align}
$Z_\delta$ becomes
\begin{align}
    \frac{Z_{\delta}}{Z_MZ_c} &= \frac{2\pi}{\sqrt{k}}\cdot\frac{2\pi}{b'}\Gamma(-2s')\left(-\frac{\mu}{2b'^2}\right)^{2s'}\frac{\Gamma(2s'+1)}{\Gamma(s'+1)^2}\nonumber\\
    &\quad \times \int \prod_{I=3}^{s'}d^2z_I\,|z_I|^2 |z_I-1|^2 \prod_{I'=2}^{s'}d^2z_{I'}\,|z_{I'}|^{2(1-k)}|z_{I'}-1|^{2(1-k)}\nonumber\\
    &\quad \times\prod_{3 \leq I < J \leq s'}|z_{IJ}|^2 \prod_{2 \leq I' < J' \leq s'}|z_{I'J'}|^2 \prod_{I,J'}|z_{IJ'}|^{2(1-k)} \nonumber\\
    &\quad \times\left( |z_3|^{2\epsilon'-4s'\delta} |z_3-1|^{-4s'\delta} \prod_{J>3}|z_{3J}|^{-4s'\delta} \prod_{J'=2}|z_{3J'}|^{-4(1-k)s'\delta} \right),\label{Z-delta,3-worldsheet}
\end{align}
In order to integrate over \( z_{I'} \), we employ the formula \cite{Halder:2022ykw}
\begin{align}
    &\frac{1}{\pi^n n!}\int \prod_{i=1}^n d^2y \left( \prod_{1\leq i<j\leq n}|y_i-y_j|^2 \right) \left( \prod_{i=1}^{n} \prod_{j=1}^{n+1}|y_i-t_j|^{2p_j} \right)\nonumber\\
 &\qquad \qquad 
 = \frac{\prod_{j=1}^{n+1} \gamma(1+p_j)}{\gamma(1+n+\sum_{j=1}^{n+1}p_j)} \prod_{1 \leq j < j' \leq n+1} |t_j-t_{j'}|^{2p_j+2p_{j'}+2}\ ,
\label{int:zIprime}
\end{align}
where $\gamma(x):=\Gamma(x)/\Gamma(1-x)$. It follows that
\begin{align}
    \frac{Z_{\delta}}{Z_MZ_c}&=\frac{ \pi}{\sin(\pi(-2s'))}\frac{2\pi}{\sqrt{k}}\cdot\frac{2\pi}{b'}\left(-\frac{\mu}{2b'^2}\right)^{2s'}\frac{\pi^{s'-1}}{s'\Gamma(s'+1)}\frac{\gamma(2-k)^{s'-1}\,\gamma(2-k-(1-k)s'\delta)}{\gamma(s'(2-k)-(1-k)s'\delta)}\nonumber\\
    &\quad\times\int\prod_{I=3}^{s'}|z_I|^{4(2-k)}|z_I-1|^{4(2-k)}\prod_{3\leq I<J\leq s'}|z_{IJ}|^{4(2-k)}\nonumber\\
&\quad \times|z_3|^{2\epsilon'-4(2-k)s'\delta}|z_3-1|^{-4(2-k)s'\delta}\prod_{I=4}^{s'}|z_{3I}|^{-4(2-k)s'\delta}\label{Z_delta,3}.
\end{align}
Using
\begin{align}
    {1\over \gamma(s'(2-k)-(1-k)s'\delta)} = (1-k)s'\delta+\mathcal{O}(\delta^2) \ ,
    \label{exp:1/gamma}
\end{align}
we find
\begin{align}
    \frac{Z_{\delta}}{Z_MZ_c}=&\,\frac{2\pi s'\delta\cdot(1-k)}{\sin(\pi(-2s'))}\frac{2\pi}{\sqrt{k}}\cdot\frac{2\pi}{b'}\left(-\frac{\mu}{2b'^2}\right)^{2s'}\frac{\pi^{s'-1}}{s'\Gamma(s'+1)}\gamma(2-k)^{s'}\nonumber\\
    &\times\int\prod_{I=3}^{s'}d^2z_I\,|z_I|^{4(2-k)}|z_I-1|^{4(2-k)}\prod_{3\leq I<J\leq s'}|z_{IJ}|^{4(2-k)}\ .
\end{align}

It is clear that $Z_\delta$ diverges when $s'=2,3,4,\cdots$. The singular behavior can be captured by analytically continuing $s'$ to a complex variable and examining the residue of $Z_\delta$:
\begin{align}
    \underset{s'\to \mathbb{Z}}{\mathrm{Res}}\left(\frac{Z_{\delta}}{Z_MZ_c}\right)
    &= -\delta\cdot a_{\rm w}(k)\frac{2\pi}{\sqrt{k}}\cdot\frac{\pi}{b'}\left(-\frac{\mu}{2b'^2}\right)^{2s'}\frac{\pi^{s'-1}}{s'\Gamma(s'+1)}\gamma(2-k)^{s'}\nonumber\\
    &\quad\times\int\prod_{I=3}^{s'}|z_I|^{4(2-k)}|z_I-1|^{4(2-k)}\prod_{3\leq I<J\leq s'}|z_{IJ}|^{4(2-k)},
\end{align}
where we defined the ``worldsheet replica factor''
\begin{align}
a_{\rm w}(k):= {2(k-1)\over k-2}\ .
\label{aw:replica}
\end{align}
As shown in \cite{Halder:2023adw}, the poles and residues are identical to those of the Liouville three-point function $C_{(b', \mu)}(b', b', b')$.
{}Following the prescription of that paper, we equate the two functions with each other up to a prefactor that is equal to $1$ when $s'$ is an integer:
\begin{align}
    \frac{Z_{\delta}}{Z_MZ_c}
    &= -(-1)^{2s'l}a_{\mathrm{w}}(k)\delta \cdot\frac{2\pi}{\sqrt{k}}\cdot \frac{2\pi}{b'} \left(-\frac{\mu}{2b'^2}\right)^{2s'} \frac{\pi^{s'-1}}{s'^2(s'-1)} \gamma(2-k)^{s'}\, (-\mu)^{2-s'} \cdot b' C_{(b', \mu)}(b', b', b') \ .
\end{align}
Here $l\in\mathbb{Z}$ is to be fixed below.
Using the so-called DOZZ formula \cite{Dorn:1994xn,Zamolodchikov:1995aa}
\begin{align}
    C_{(b', \mu)}(\alpha_1, \alpha_2, \alpha_3) & = \left[\pi \mu \gamma(b'^2) b'^{2-2b'^2}\right]^{\frac{Q' - \sum_k \alpha_k}{b'}} \frac{\Upsilon'_{b'}(0)\, \prod_{k=1}^3 \!\Upsilon_{b'}(2\alpha_k)}{\Upsilon_{b'}(\sum_k \alpha_k - Q') \,\prod_{k=1}^3\! \Upsilon_{b'}(\sum_j \alpha_j - 2\alpha_k)}
\end{align}
with $Q' := b' + b'^{-1}$, we find
\begin{align}
    C_{(b', \mu)}(b', b', b') &= \left[\pi \mu \gamma(b'^2) b'^{2-2b'^2}\right]^{s'-2} \frac{\Upsilon'_{b'}(0)\, \Upsilon_{b'}(2b')^3}{\Upsilon_{b'}(-(s'-2)b') \,\Upsilon_{b'}(-b')^3}.
\end{align}
It is found that $\Upsilon_{b'}(-(s'-2)b')$ is not well-defined. The paper \cite{Halder:2023adw} proposes to define it by a limit
\begin{align}
    \lim_{\epsilon\to 0}\Upsilon_{b'}(-(s'+\epsilon-2)b') \ .\nonumber
\end{align}
Using this prescription, we obtain
\begin{align}
    \frac{Z_{\delta}}{Z_MZ_c}
    &= -(-1)^{2s'(l+1)}a_{\mathrm{w}}(k)\delta \cdot\frac{2\pi}{\sqrt{k}}\cdot \mu^{\frac{2}{k-2}} \frac{4^{\frac{1}{2-k}} (k-3) (k-2)^{\frac{3k+2}{4-2k}} \pi^{\frac{2}{k-2} - 2} \Gamma\left(\frac{1}{2-k}\right)}{\Gamma\left(1 + \frac{1}{k-2}\right)}. \label{partition function on worldsheet}
\end{align}

\subsection{String Entanglement Entropy for Sine-Liouville String}

The string entanglement entropy is computed from the $n\to 1$ limit of the R\'enyi entropy $\mathcal{S}_n$ in~\eqref{string_Renyi_multi-particle} as
\begin{align}
       \mathcal{S}_{\mathrm{EE}} 
       =\lim_{n\to 1}\mathcal{S}_n= \lim_{\delta\to 0}\partial_\delta \bigg(  \Tr_L[\rho_L^{1-\delta}]  - (1-\delta) \Tr_L[\rho_L] \bigg) .
\end{align}
As discussed in \eqref{order_delta_contrib_to_tr_rho_n}, at $\mathcal{O}(\delta)$, $\Tr_L[\rho_L^{1-\delta}]$ consists of two parts: the vertex operator contribution which, is nothing but $Z_\delta$ worked out in the previous section (see \eqref{Tr_Zdelta}), and the replica contribution. Since we currently lack computational tools to evaluate the replica contribution, we focus on the vertex operator contribution.  We have
\begin{align}
    \mathcal{S}_{\mathrm{EE}} 
     = \lim_{\delta\to 0}\partial_\delta \bigg(  Z_\delta - (1-\delta) Z_{\delta=0} \bigg)+ S_{\mathrm{rep}} \ ,
\end{align}
where $S_{\mathrm{rep}}$ is the replica contribution to the string entanglement entropy. Using \eqref{partition function on worldsheet}, we find
\begin{align}
    \mathcal{S}_{\mathrm{EE}} &= -(-1)^{2s'(l+1)}\frac{1}{g_s^2} Z_c Z_M a_{\mathrm{w}}(k) \cdot \frac{2\pi}{\sqrt{k}} \cdot \mu^{\frac{2}{k-2}} \frac{4^{\frac{1}{2-k}} (k-3)(k-2)^{\frac{3k+2}{4-2k}} \pi^{\frac{2}{k-2}-2} \Gamma\left(\frac{1}{2-k}\right)}{\Gamma\left(1+\frac{1}{k-2}\right)} + S_{\mathrm{rep}}\ ,
    \label{moyb22May25}
\end{align}
where we inserted the factor $1/g_s^2$ for the genus zero worldsheet (see \eqref{Phi_sL}).  Substituting the cosmological constant $\mu$ with the explicit value \cite{Halder:2024gwe}
\begin{align}
\mu = 2(k-2)^2\pi^{-\frac{k}{2}} \left( -\frac{\Gamma\left(\frac{1}{k-2}\right)}{\Gamma\left(\frac{1}{2-k}\right)} \right)^{\frac{k-2}{2}}\ ,
\end{align}
we obtain
\begin{align}
    \mathcal{S}_{\mathrm{EE}} &= \frac{1}{g_s^2} Z_c Z_M \frac{2}{\sqrt{k}} \frac{2(k-1)}{k-2} \frac{1}{\pi^2} \left( \sqrt{k-2} - \frac{1}{\sqrt{k-2}} \right) + S_{\mathrm{rep}}\ , \label{final result}
\end{align}
where we set $l=-1$ to remove a complex phase. 

We claim that this gives the thermal entropy of a two-dimensional black hole that is valid for any finite $k$ with the $\alpha'$ corrections fully incorporated. Unfortunately, we are unable to verify it because $S_{\mathrm{rep}}$ remains undetermined. Instead, we compare this result with the spacetime thermal entropy of the cigar-shaped Euclidean black hole computed in \cite{Halder:2024gwe} and reviewed in section
\ref{sec:review_of_Halder-Jafferis}:
\begin{align}
    \mathcal{S}_{\mathrm{thermal}} &:= \frac{1}{g_s^2} Z_c Z_M \cdot \frac{2\pi}{\sqrt{k}} \cdot  \frac{2k}{k-2} \cdot \frac{1}{\pi^3} \left( \sqrt{k-2} - \frac{1}{\sqrt{k-2}} \right)\ ,\label{2-dim_thermal_entropy}
\end{align}
where $\tfrac{2\pi}{\sqrt{k}}$ comes from the $\hat\theta$ zero mode integral, and the factor
\begin{align}
 a_{\rm s}(k)\equiv\frac{2k}{k-2}
 \label{eq:a_t_def}
\end{align}
is called the replica factor in \cite{Halder:2024gwe}, which we here refer to as the spacetime replica factor, because $ \mathcal{S}_{\mathrm{thermal}}$ is computed by means of a spacetime replica trick following \cite{Lewkowycz:2013nqa}. In contrast, $\mathcal{S}_\mathrm{EE}$ is the string entanglement entropy by definition. We note that the vertex operator contribution takes the same form as the thermal entropy with the replica factor replaced by worldsheet replica factor $a_{\mathrm{w}}(k)$ defined in \eqref{aw:replica}.

Equating the two entanglement entropies, we conjecture that \( S_\mathrm{rep} \) is given by
\begin{align}
    \mathcal{S}_{\mathrm{rep}} 
    &= \frac{1}{g_s^2} Z_c Z_M \cdot \frac{4}{\pi^2\sqrt{k}(k-2)}\left( \sqrt{k-2} - \frac{1}{\sqrt{k-2}} \right)\ .
    \label{eq:S_rep_2d}
\end{align}
{}For $k>3$, we find that $S_{\mathrm{rep}}$ is positive. This provides us with a consistency check for interpreting 
$S_{\mathrm{rep}}$ to be part of the string entanglement entropy.
Furthermore, we compute the ratio
\begin{align}
    \frac{\mathcal{S}_{\rm EE}-\mathcal{S}_{\mathrm{rep}}}{\mathcal{S}_{\mathrm{thermal}}}=\frac{k-1}{k}\ .\label{ratio_of_Sth_and_S_EE_in_2dim}
\end{align}
This shows that in the large $k$ limit, the vertex operator contribution dominates $\mathcal{S}_{\mathrm{EE}}$ compared with $\mathcal{S}_{\mathrm{rep}}$.

Note that $S_{\mathrm{rep}}$ should be regarded as a classical effect.  This is because $S_{\mathrm{rep}}$ is obtained by taking the $\delta \to 0$ limit in the worldsheet Renyi entropy, even though it is generically defined on a higher-genus replica worldsheet.  In this limit, $S_{\mathrm{rep}}$ goes as $g_s^{-2}$, indicating that this is a classical contribution. Then, all the quantities involved in the comparison of
  the three entropies are classical, being proportional to $g_s^{-2}$.

\subsection{Generalization to Three-Dimensional Black Holes}
\label{Generalization to Three-Dimensional Black Holes}
In this subsection, we work with a 3d extension
of the FZZ duality \cite{Jafferis:2021ywg} (see also \cite{Halder:2022ykw}),  which states that the Euclidean BTZ black hole background is dual to a 3d sine-Liouville background. 
We attempt to compute the entropy of the BTZ black hole in terms of the string entanglement entropy for a 3d sine-Liouville string theory, which
is constructed from the extended sine-Liouville together with internal and ghost CFTs.

The metric of the Euclidean BTZ black hole geometry at temperature $T$ in the large $k$ limit is given by
\begin{align}
ds^2=l_{\mathrm{AdS}}^2(d\hat{r}^2+\cosh^2\hat{r}\,d\hat{\theta}^2+\sinh^2\hat{r}\,d\hat{\xi}^2) \ ,
\end{align}
with 
$\hat{\theta} \sim \hat{\theta} + 4\pi^2 T l_\mathrm{AdS}$ and $\hat{\xi} \sim \hat{\xi} + 2\pi$.
$\xi$ is the thermal coordinate of the Euclidean black hole, which smoothly collapses to zero at the tip of the cigar, $\hat{r}=0$.
The winding number of strings wrapped around the $\xi$ direction is not conserved.

The 3d extension of the FZZ duality says that the CFT on the Euclidean BTZ black hole is dual to an extended sine-Liouville CFT with the action given by \begin{align}
    S_{\rm EBTZ}&=\frac{1}{2\pi}\int d^2\sigma\,2\left(\partial\varphi\Bar{\partial}\varphi+\frac{1}{4b'}\sqrt{h}\mathcal{R}\varphi+\beta\Bar{\partial}\gamma+\Bar{\beta}\partial\Bar{\gamma}+4\pi\lambda(W^++W^-)\right)\ ,\\
    W_{\pm}&=e^{\pm i\frac{\sqrt{k}}{4}(\gamma-\Bar{\gamma})}e^{\pm i\sqrt{k}(\int^{z}\beta dz'-\int^{\Bar{z}}\Bar{\beta}d\Bar{z}')}e^{b'\varphi}\ .
\end{align}
The target space coordinates in the sine-Liouville CFT are mapped to those in the BTZ CFT by $\gamma=\sqrt{k}(\hat{\xi}+i\hat{\theta})$ and $\varphi=-\sqrt{k}\hat{r}$; see also \cite{Halder:2023adw}.

\subsubsection*{Spacetime thermal entropy}

We first derive the spacetime thermal entropy of the BTZ black hole using the 3d sine-Liouville string theory description. This is a 3d version of the 2d computation
in \cite{Halder:2024gwe} that we reviewed in section 
\ref{sec:review_of_Halder-Jafferis}, and is based on the idea to insert
winding string vertex operators on the dual sine-Liouville side so that a conical deficit is introduced at the tip of a cigar.
The first computation for the case of the 3d sine-Liouville background was made in \cite{Halder:2023adw},
where the inserted string vertex operators were chosen to be marginal, $h=\bar{h}=1$. However, in \cite{Halder:2024gwe}, it was argued that,
in order to reproduce the known $\alpha'$ correction to the thermal 2d black hole entropy, the vertex operators must be allowed to be non-marginal with an anomalous dimension of $\mathcal{O}(\delta)$.
Therefore, even for computing the thermal entropy for the 3d setting, it is more appropriate to modify the vertex operators by allowing them to be non-marginal. This is what we do in the following.  So, we can say that the following is a 3d version of the 2d computation in \cite{Halder:2024gwe}, making the result in~\cite{Halder:2023adw} more precise.

As in the 2d case reviewed in Sec.~\ref{sec:review_of_Halder-Jafferis}, we introduce a conical singularity by modifying the vertex operators
\begin{align}
    W_{\pm}\xrightarrow{k\to k(1-2\delta)} W^\delta_{\pm}:=e^{\pm i\frac{\sqrt{k}}{4}(1-\delta)(\gamma-\Bar{\gamma})}e^{\pm i\sqrt{k}(1-\delta)(\int^{z}\beta dz'-\int^{\Bar{z}}\Bar{\beta}d\Bar{z}')}e^{b'\left(1-\frac{\delta}{1-\frac{2}{k}}\right)\varphi}
\end{align}
while keeping the periodicity of $\hat{\theta}$ and $\hat{\xi}$ unchanged. 
This effectively changes the geodesic circumference of the $\hat{\theta}$ circle.
We use the symbol $Z_{M'}$ to denote the partition function of the internal spacetime. As in the two-dimensional case in Sec.~\ref{Evaluation of the String Entanglement Entropy}, the regularized partition function in the three-dimensional analog of the sine-Liouville string is given by
\begin{align}
    Z^{\rm th,3d}_{\delta} &:= 8\pi^3(Tl_\mathrm{AdS})\cdot Z_\mathrm{M'} Z_c \frac{2\pi}{b'} \Gamma(-2s') \left(\frac{\mu}{2b'^2}\right)^{2s'} \frac{\Gamma(2s'+1)}{\Gamma(s'+1)^2} \int \mathcal{D}\hat{\varphi}' \mathcal{D}^2\hat{\gamma}'\mathcal{D}^2\hat{\beta}'\mathcal{D} e^{-S_{\mathrm{EBTZ}}|_{\lambda=0}} \nonumber \\
    &\quad \times W_+(0)\, W_+(1)\, W_-(\infty) \int d^2z_3 \, W^{2s'\delta}_+ \prod_{j=4}^{s'} \int d^2z_j \, W_+ \prod_{k=2}^{s'} \int d^2z_k \, W_-, \label{z_delta_3}
\end{align}
here the factor $8\pi^3(Tl_\mathrm{AdS})$ comes from the zero mode contribution of $\hat{\theta}$ and $\hat{\xi}$, and $s'=1/(k-2)$. Using the OPEs
\begin{align}
    &W^\delta_\pm(z,\Bar{z})W_{\pm}(w,\Bar{w})\sim|z-w|^{2}:W_\pm W_{\pm}(w,\Bar{w}):~,\nonumber\\
    &W^\delta_{\pm}(z,\Bar{z})W_{\mp}(w,\Bar{w})\sim|z-w|^{2(1-k+k\delta )}:W_{\pm}W_{\mp}(w,\Bar{w}):~,
\end{align}
the string partition function becomes
\begin{align}
    Z^{\rm th,3d}_{\delta} &= Z_c Z_{M'} \cdot 8\pi^3(Tl_\mathrm{AdS}) \cdot\frac{2\pi}{b'}\Gamma(-2s')\left(-\frac{\mu}{2b'^2}\right)^{2s'}\frac{\Gamma(2s'+1)}{\Gamma^2(s'+1)}\nonumber\\
    &\quad \times \int \prod_{I=3}^{s'}|z_I|^2 |z_I-1|^2 \prod_{I'=2}^{s'}|z_{I'}|^{2(1-k)}|z_{I'}-1|^{2(1-k)}\nonumber\\
    &\quad \times\prod_{3 \leq I < J \leq s'}|z_{IJ}|^2 \prod_{2 \leq I' < J' \leq s'}|z_{I'J'}|^2 \prod_{I,J'}|z_{IJ'}|^{2(1-k)} \nonumber\\
    &\quad \times\left( |z_3|^{2s'(6k-2+\frac{4k}{k-2})\delta} |z_3-1|^{2s'k\delta} \prod_{J>3}|z_{3J}|^{2s'k\delta} \prod_{J'=2}|z_{3J'}|^{2a_\mathrm{s}(k)\delta} \right),
\end{align}
Following the same procedures as in section \ref{Evaluation of the String Entanglement Entropy}, we find
\begin{align}
    Z^{\rm th,3d}_{\delta} &=-a_{\mathrm{s}}(k)\delta\cdot Z_c Z_{M'} \cdot 8\pi^3(Tl_\mathrm{AdS}) \cdot \frac{1}{\pi^3} \left( \sqrt{k-2} - \frac{1}{\sqrt{k-2}} \right)\ ,
\end{align}
where the spacetime replica factor $a_{\mathrm{s}}(k)=2ks'$ is identical with that in two dimensions. From this, \eqref{sine_Liouville_nth_renyi_entropy}, and the cosmological constant given in \cite{Halder:2024gwe}, we obtain the thermal black hole entropy
\begin{align}
    \mathcal{S}_{\mathrm{thermal}}^{\mathrm{{3d}}} = \frac{1}{g_s^2} Z_c Z^{M'} \cdot 8\pi^3(Tl_\mathrm{AdS}) \cdot a_{\mathrm{s}}(k) \cdot \frac{1}{\pi^3} \left( \sqrt{k-2} - \frac{1}{\sqrt{k-2}} \right).
\end{align}
The only difference from the two-dimensional case \eqref{2-dim_thermal_entropy} is $Z_{M'}$ and the factor $8\pi^3(Tl_\mathrm{AdS})$.

\subsubsection*{String entanglement entropy}

We now turn to the string entanglement entropy, focusing on the vertex operator contribution. We modify the vertex operator as
\begin{align}
    W^\delta_{\pm}\to W^\delta_{\pm}:=e^{\pm i\frac{\sqrt{k}}{4}(1-\delta)(\gamma-\Bar{\gamma})}e^{\pm i\sqrt{k}(1-\delta)(\int^{z}\beta dz'-\int^{\Bar{z}}\Bar{\beta}d\Bar{z}')}e^{b'(1-\delta)\varphi}
\end{align}
so that the boundary condition does not depend on the replica number of the worldsheet (see \ref{app:neutrality condition}) with the periodicity of $\hat{\theta}$ and $\hat{\xi}$ remaining unchanged. The OPEs of the modified vertex operator are
\begin{align}
    &W^\delta_\pm(z,\Bar{z})W_{\pm}(w,\Bar{w})\sim|z-w|^{2(1-\delta)}:W_\pm W_{\pm}(w,\Bar{w}):~,\nonumber\\
    &W^\delta_{\pm}(z,\Bar{z})W_{\mp}(w,\Bar{w})\sim|z-w|^{2(1-k)(1-\delta)}:W_{\pm}W_{\mp}(w,\Bar{w}):~.
\end{align}
Although we are working in the three-dimensional case, the singular structures \( |z-w|^{2(1-\delta)} \) and \( |z-w|^{2(1-k)(1-\delta)} \) have the same form as those in the two-dimensional case, \eqref{ope-delta}. Using these OPEs, we obtain the string entanglement entropy as follows:
\begin{align}
    \mathcal{S}_{\mathrm{EE}}^{\mathrm{3d}} = \frac{1}{g_s^2} \cdot a_{\mathrm{w}}(k) \cdot Z_c Z_{M'} \cdot 8\pi^3(Tl_\mathrm{AdS}) \cdot \frac{1}{\pi^3} \left( \sqrt{k-2} - \frac{1}{\sqrt{k-2}} \right)+\mathcal{S}_{\mathrm{rep}}^{\mathrm{3d}}\ ,
\end{align}
where $\mathcal{S}_{\mathrm{rep,3d}}$ is the replica contribution to the entanglement entropy. Note that the worldsheet replica factor is common both in 2d and 3d.

\subsubsection*{Comparison}

We compare the thermal entropy with the vertex operator contribution of the three-dimensional sine-Liouville string. The only difference between them is $a(k)$ and $a_{\mathrm{w}}(k)$, then the ratio of them is
\begin{align}
    \frac{\mathcal{S}_{\mathrm{EE}}^{\mathrm{3d}}-\mathcal{S}_{\mathrm{rep}}^{\mathrm{3d}}}{\mathcal{S}_{\mathrm{thermal}}^{\mathrm{3d}}}&=\frac{a_{\mathrm{w}}(k)}{a_{\mathrm{s}}(k)}=\frac{k-1}{k}.
\end{align}
This result is the same as that in the two-dimensional case \eqref{ratio_of_Sth_and_S_EE_in_2dim}. 

\section{Summary and Outlook}

In this paper, we discussed how the 2d and 3d black-hole thermal entropies can be accounted for by the entanglement entropy of folded strings. As found in \cite{Jafferis:2021ywg}, the folded strings arise in the sine-Liouville string theory from condensation of winding string vertex operators $W_{\pm}$.   The string entanglement entropy is computed using a worldsheet replica trick:  We define the reduced density matrix of a pair of folded string groups in terms of a worldsheet path integral with branch cuts running between $W_{\pm}$ on the Riemann sphere.
The R\'enyi entropy associated with the density matrix is written in terms of the path integral on a replica worldsheet that is constructed by gluing multiple Riemann sheets along the branch cuts. This is implemented by modifying winding string vertex operators in an appropriate manner, which plays the role of inserting twist operators at the branch points and the emergence of folded strings simultaneously. This is clearly distinct from the standard replica method. In fact, it is found that the entanglement entropy consists of two contributions, one from the string partition function on a sphere with modified vertex operators (``vertex operator contribution'') and the other from a worldsheet path integral on a higher-genus Riemann surface (``replica contribution''). The former can be computed using the known results for the Liouville three-point functions.  On the other hand, we currently lack computational tools to directly evaluate the latter; we instead indirectly infer its form by demanding that the string entanglement entropy coincide with the thermal black-hole entropy computed in 2d \cite{Halder:2024gwe} and its straightforward extension to the 3d black hole. The resultant expression is consistent in that it is always positive; this can be regarded as support for the interpretation of the string entanglement entropy as a measure of entanglement of folded string pairs.

Several questions remain to be clarified for a deeper understanding of the nature of string entanglement entropy. One of the most important ones is to directly compute the replica contribution to the entanglement entropy. To this end, one has to consider the sine-Liouville CFT on a Riemann surface of genus $g$ and make an analytic continuation of $g$ to $-(s^\prime-1)\delta$.  One possible way to address this problem would be to start with the case of a free boson with condensation of winding strings in order to gain insight into the sine-Liouville CFT case.

As found in this paper, the string entanglement entropy for both the 2d and 3d black holes takes a universal form in that the only difference is the partition functions of the internal CFTs.
It would be interesting to understand why and examine if the universal behavior persists for higher-dimensional black holes. 

It was observed that, for $k\gg 1$, the thermal black-hole entropy is dominated by the vertex operator contribution. It would be interesting to show that the replica contribution is less dominant in the large $k$ limit by analyzing it directly without using the requirement $\mathcal{S}_{\mathrm{EE}}=\mathcal{S}_{\mathrm{thermal}}$.

Comparison with the result of \cite{Halder:2024gwe} suggests that the replica contribution is quite universal and independent of the detail of the internal CFT (see \eqref{eq:S_rep_2d}). This is quite non-trivial because, in order to evaluate the replica contribution, we must compute the entanglement entropy not just for the sine-Liouville and ghost CFTs but also the internal CFT.  Note that the total central charge of the worldsheet CFT being zero does not imply that the total entanglement entropy vanishes.
This presumably means that there is a physical reason for the replica contribution being simple.  One possible scenario is the following. Unlike the standard CFTs, the string worldsheet CFT should be integrated over the moduli space of the Riemann surface.  It is logically possible that the string entanglement entropy can be written as a total derivative with respect to the worldsheet moduli, and integrating the result over the moduli space localizes to the boundary of the moduli space.  It is conceivable that such a mechanism relates the result at general genus $g$ to that of lower genus where results are simple.

We remark that, although our prescription for computing string entanglement entropy gave quite reasonable results, its validity is not fully established. One of the most important problems to understand is how to handle the moduli of the replica worldsheet.\footnote{We would like to thank Edward Witten for pointing this out to us.} It takes the form of an $n$-sheeted branched cover, which is a Riemann surface of genus $g'=(s'-1)(n-1)$. When computing string R\'enyi entropy, however, we integrated only over the $2s'-3$ moduli of the Riemann surface, simply neglecting the remaining moduli. To see why the moduli integral problem is important to clarify, note that we defined string entanglement entropy by dividing the worldsheet Hilbert space into left and right Hilbert spaces with the worldsheet moduli fixed, and then integrating the string entanglement entropy over the worldsheet moduli. Although this is a completely natural prescription for computing CFT entanglement entropy using the replica method, in the string theory context, this might imply that the string entanglement entropy before the integral is unphysical because of the contributions from off-shell, unphysical string modes. This is in analogy with the fact that string amplitudes become physically meaningful only after integration over the worldsheet moduli, because unphysical intermediate modes disappear after the integration. 
Therefore, it is desirable to prove that the moduli integral makes the string entanglement entropy consistent, by confirming that it eliminates the contributions from unphysical modes; we currently do not know how to carry out this proof.
However, the fact that our procedure led to reasonable expressions for entanglement entropy supports that we are on the right track.   This issue deserves further investigation.

As a potential obstruction to this, divergences associated with tadpoles could arise when integrating over worldsheet moduli in computing $S_{\mathrm{rep}}$, much as those relevant for the Fischler-Susskind mechanism \cite{Fischler:1986tb,Fischler:1986ci}.
For canceling them, the off-shell string formulation in \cite{Ahmadain:2022eso} is expected to be useful for finding out an appropriate shift of the sine-Liouville string background. 


Related to the problem of the worldsheet moduli, furthermore, it would be very interesting to define a stringy area operator as was done in \cite{Halder:2024gwe}.  This would be possible only after gaining a better understanding of the mathematical structure of $S_{\mathrm{rep}}$. 
We leave this as an interesting future study.

\section*{Acknowledgments}

We would like to thank Yasuaki Hikida for useful discussions. 
The work of SM was financially supported by JST SPRING, Grant Number JPMJSP2125.  SM would like to take this opportunity to thank the “THERS Make New Standards Program for the Next Generation Researchers.” 
The work of MS was supported in part by MEXT KAKENHI Grant Numbers 21H05184 and 24K00626.

\appendix

\section{Replica trick with vertex insertions}
\label{app:replica_with_vtx}

\subsection{Boundary conditions and modification of vertex operators}
\label{app:boundary conditions}

As we discussed in section \ref{ss:folded strings and string entanglement entropy}, when we do the replica trick by going to the $n$-cover,  we need to modify the vertex operators $W_\pm$ inserted at the branch points as in \eqref{def_Wn}.  Here we explain how this comes about.

In the standard replica trick, no operators are inserted at the branch points but, in our situation, we have insertions there.  To understand how to go to the covering space in this situation, it is useful to look at the insertions using angular quantization \cite{Agia:2022srj} in which one takes the angle around the insertion as time.\footnote{In general, this angular time cannot be defined away from the insertion, but we are only interested in the region near the insertion. }  In angular quantization, one replaces a vertex operator with a boundary condition imposed at small distance
$\epsilon$ from the position of the vertex operator, and that boundary
condition defines a Hilbert space on a constant angle slice and an
angular Hamiltonian with which to evolve in the angular direction.  In
the original space, the angular direction is $2\pi$ periodic and we
evolve with the angular Hamiltonian for angular time $2\pi$.  Going to
the $n$-cover means to use the same boundary condition and evolve with
the same angular Hamiltonian for angular time $2\pi n$.

For example, in the free $X$ CFT, assume that the vertex operator $e^{ikX}$
is inserted at $z=0$ in the original space. When there are multiple
insertions, we should regard this as a local description near one of
them.  The OPE of the $X$ field and the inserted operator is
\begin{align}
 \partial_z X(z) \, [e^{ikX}(0,0)]_1 \sim -{ik\over 2z}\,[e^{ikX}(0,0)]_1,\label{llco8Apr24}
\end{align}
where $[\cO]_1$ means the normal-ordered operator using the free
propagator\footnote{When there are multiple insertions, this is a local expression.}
\begin{align}
 \Delta_1(z,z')=-{1\over 2}\log|z-z'|^2.
\end{align}
  So,
inserting the operator $[e^{ikX}]_1$ is equivalent to imposing at
$|z|=\epsilon\ll 1$ the boundary condition
\begin{align}
 \partial_z X(z) \stackrel{\rm b.c.}{=} -{ik\over 2z},\label{lsqe8Apr24}
\end{align}
where ``b.c.''~means that we impose this as a boundary condition, 
and the anti-holomorphic counterpart.\footnote{More precisely, there is a choice
between Neumann-like and Dirichlet-like boundary conditions
\cite{Agia:2022srj}}  When we go to the covering space by extending the
range of $\arg z$ to $[0,2\pi n)$, we have to find a vertex operator
inserted at $z=0$ that has the same OPE with $X$ as \eqref{llco8Apr24}
on all the $n$ sheets.  When we define normal-ordered operators on the
$n$-cover, we must use for subtraction the correlator on the $n$-cover,
\begin{align}
 X(z,\zb)X(z',\zb')\sim -{1\over 2}\log|z^{1/n}-z'^{1/n}|^2=: \Delta_n(z,z').
\end{align}
We denote by $[\cO]_n$ the normal-ordered operator defined by
subtraction of $\Delta_n$.  Using this propagator, we can see that
\begin{align}
 \partial_z X(z) \, [e^{inkX}(0,0)]_n 
 \sim  ink\partial_z \Delta_n(z,0)\,[e^{inkX}(0,0)]_n
 =  -{ik\over 2z}\,[e^{inkX}(0,0)]_n.
\end{align}
Namely, it is $[e^{inkX}(0,0)]_n$ that leads to the same boundary
condition \eqref{lsqe8Apr24} and therefore to the same angular
Hamiltonian on the $n$-sheeted cover.
In summary, the replica trick with the vertex operator $[e^{ikX}]_1$
inserted at a branch point means to go to an $n$-cover with the vertex
operator $[e^{inkX}]_n$ inserted at the branch points.

Now assume that $X=X_L(z)+X_R(\zb)$ is compact with periodicity
$2\pi R$.  If we have a vertex operator $[e^{iR(X_L-X_R)}]_1$ at a branch
point, in the base space we have the boundary condition
\begin{align}
 \partial_z X_L  \stackrel{\rm b.c.}{=} -{iR\over 2z},\qquad
 \partial_\zb X_R\stackrel{\rm b.c.}{=} {iR\over 2\zb}\label{mpkh8Apr24}
\end{align}
at $|z|=\epsilon$.  This implies that, when we go around the insertion as
$z\to e^{2\pi i}z$, the target space $X=X_L+X_R$ changes as $X\to X+2\pi
R$, meaning that the string wraps once around the $X$ circle. If we go
to an $n$-cover, as we discussed above, we must insert
$[e^{inR(X_L-X_R)}]_n$ which induces the same boundary condition
\eqref{mpkh8Apr24} on all the sheets.  As we go around the insertion in the
covering space as $z\to e^{2\pi i n}z$, $X$ changes as $X\to X+2\pi R n$,
meaning that the string wraps $n$ times around the $X$ circle.

In the neighborhood of the insertion, we can define a local coordinate
$w=z^{1/n}$ in terms of which the angular periodicity is $\arg
w\cong\arg w+2\pi$ and there is no conical singularity.  In the $w$
coordinate, the vertex operator $[e^{inkX}(z)]_n$ is simply
$[e^{inkX}(w)]_1$ because $\Delta_n(z,z')=\Delta_1(w,w')$.

\bigskip
In the situation in the main text, because the linear dilaton CFT has the same OPE as the $X$ CFT, when we go to the $n$-cover, we need to modify the vertex operator $W_\pm$ as in~\eqref{def_Wn}.

\subsection{The neutrality condition}
\label{app:neutrality condition}

In the linear dilation CFT, the charge of the vertex operators is constrained by the neutrality condition that depends on the genus of the worldsheet.
Here we show that modifying the vertex operator as in \eqref{def_Wn} as we go to the $n$-cover is consistent with the neutrality condition.

The sine-Liouville theory \eqref{sine-Liouville action}  can equivalently be formulated using the following action:
\begin{align}
    S_{\mathrm{sL}} &= S_0+S_1+S_{\rm int},\label{S_sL_alternative}
\end{align}
where    
\begin{subequations}
\begin{align}
    S_0&=
    \frac{1}{4\pi} \int_\Sigma d^2\sigma \sqrt{h} ( (\nabla \hat{r})^2 + (\nabla \hat{\theta})^2),\\
    S_1&= -Q\left(\frac{1}{4\pi} \int_\Sigma d^2\sigma \sqrt{h} \,R[h]\hat{r}
        +{1\over 2\pi}\int_{\partial\Sigma} d\sigma\sqrt{\tilde{h}} \,K\hat{r}\right),\\
    S_{\rm int}
     &=\lambda\int_\Sigma d^2\sigma\sqrt{h}(W_+^{\rm bare} + W_-^{\rm bare}).
\end{align}
\end{subequations}
Here, $\tilde{h}$ is the induced metric on the boundary $\partial\Sigma$ of the worldsheet $\Sigma$, and $K$ is the extrinsic curvature of $\partial\Sigma$.
The curvature coupling $S_1$ contains the contribution from the boundary as well.  If $\hat{r}$ is constant, $S_0$ is proportional to the Euler characteristic of the worldsheet,
\begin{align}
    \chi={1\over 4\pi}\int_\Sigma d^2\sigma \sqrt{h}\, R[h]
        +{1\over 2\pi}\int_{\partial\Sigma} d\sigma\sqrt{\tilde{h}}\, K.
\end{align}
The ``bare'' vertex operators are defined by
\begin{align}
    W_\pm^{\rm bare} := e^{-(2b_{\rm sL}-{1\over 2b_{\rm sL}})\hat{r}}e^{\pm i\sqrt{k}(\hat{\theta}_L-\hat{\theta}_R)}.
\end{align}

Path integral with the action \eqref{S_sL_alternative} makes sense only after bringing down $W_\pm^{\rm bare}$ from the exponential as we did in the main text around \eqref{Z_sl_W_brought_down}.  Then the partition function is
\begin{align}
  Z_{\rm sL} \propto \int \cD \hat{r}' \cD \hat{\theta}\, e^{-S_0-S_1}\left(\int d^2z(W_+^{\rm bare} + W_-^{\rm bare})\right)^{2s'}.
  \label{llvw8May24}
\end{align}
For regularization, we need to cut out a small disk of radius $\epsilon$ around each of $W^\pm_{\rm bare}$ insertions.  Via the curvature coupling $S_1$, each of these disks contributes a factor $e^{-S_1} \sim  e^{Q\chi \hat{r}}=e^{-Q\hat{r}}=e^{-(1/2b_{\rm sL})\hat{r}}$, because we can regard $\hat{r}$ as constant on the small disk and because removing a disk from the worldsheet changes $\chi$ by $-1$. Dressed with this contribution, $W_\pm ^{\rm bare}$ becomes  $W_\pm$, and \eqref{llvw8May24} reduces to \eqref{Z_sl_W_brought_down}.

The neutrality condition for the $\hat{r}$ charge requires that
\begin{align}
 2Q-4s'b_{\rm sL}=0,\qquad s'={Q\over 2b_{\rm sL}^2}={1\over k-2},
 \label{lpku8May24}
\end{align}
where the first term comes from $S_1$ because the bulk of the worldsheet is an $S^2$ (or from the $\delta$-function curvature source at infinity), while the second term is from $2s'$ insertions of $W_\pm$.

When we do the replica trick and go to an $n$-sheeted cover, we replace $W_\pm$ by $W_\pm^{(n)}\propto e^{-nb_{\rm sL}\hat{r}}$.  Or, in the present formulation, we replace $W_\pm^{\rm bare}\propto e^{-(2b_{\rm sL}-1/2b_{\rm sL})\hat{r}}$ by $W_\pm^{{\rm bare},(n)}\propto e^{-n(2b_{\rm sL}-1/2b_{\rm sL})\hat{r}}$, in order to keep the boundary condition on the disk unchanged.  Because the circumference of the cut-out disk is now $n$ times as long, its contribution to the curvature term also changes as $e^{(1/2b_{\rm sL})\hat{r}}\to e^{(n/2b_{\rm sL})\hat{r}}$.  When multiplied by this, $W_\pm^{{\rm bare},(n)}$ becomes $W_\pm^{(n)}$. The neutrality condition now reads
\begin{align}
 -2nQ-4ns'b_{\rm sL}=0,
 \label{erbc26Mar25}
\end{align}
including contributions from the $n$ sheets. The first term is from
the curvature of the $n$ copies of $S^2$ (or from $n$ $\delta$-function
curvature sources at the infinity of the $n$ sheets) and the second term
is from $2s'$ $W_\pm^{(n)}$ insertions (or equivalently, from $2s'$
$W_\pm^{{\rm bare},(n)}$ insertions and the extrinsic curvature from $2s'$
disks of circumference $2\pi n \epsilon$).
Eq.~\eqref{erbc26Mar25} and gives the same $s'$ in \eqref{lpku8May24}.
So, modifying the vertex operator as in \eqref{def_Wn} as we go to the $n$-cover is consistent with the neutrality condition.

\section{String theory with non-marginal vertex operators}
\label{app:String theory with non-marginal vertex operators}
\subsection{Conformal Killing group and Faddeev-Popov procedure on a sphere}

Consider an $SL(2,\mathbb{C})$ transformation of three reference points
$z_{1,2,3}$ on a Riemann sphere
\begin{align}
  z_i^\prime=\frac{pz_i+q}{rz_i+s} \ ,~~~
ps-qr=1 \ .\nonumber
\end{align}
An infinitesimal $SL(2,\mathbb{C})$ transformation about $z_i^\prime$ reads
\begin{align}
  d  z_i^\prime
=
\frac{(rz_i+s)(d  p\, z_i+d  q)-(pz_i+q)(d  r\, z_i+d  s)}
{(rz_i+s)^2} \ .\nonumber
\end{align}
We compute the wedge product $d  z_1^\prime\wedge  d  z_2^\prime\wedge  d  z_3^\prime$.
Using 
\begin{align}
  d  p\wedge d  q\wedge d  s
&=\frac{q}{p}d  p\wedge d  q\wedge d  r \ ,
\nonumber\\
  d  p\wedge d  r\wedge d  s
&=-\frac{r}{p}d  p\wedge d  q\wedge d  r \ ,
\nonumber\\
  d  q\wedge d  r\wedge d  s
&=-\frac{s}{p}d  p\wedge d  q\wedge d  r \ ,
\nonumber
\end{align}
a bit lengthy computation gives
\begin{align}
  d  z_1^\prime\wedge  d  z_2^\prime\wedge  d  z_3^\prime
=
-  \frac{2}{p}
\frac{z_{12}z_{13}z_{23}}{(rz_1+s)^2(rz_2+s)^2(rz_3+s)^2}
d  p\wedge d  q\wedge d  r \ .
\nonumber
\end{align}
Here, $z_{ij}=z_i-z_j$.

Let $V_i(z)$ be a conformal primary field  with
conformal dimension $1$. This transforms under the $SL(2,\mathbb{C})$
as
\begin{align}
  V_i'(z^\prime)
=
\frac{\partial z}{\partial z^\prime}
V_i(z,\bar{z})
=
(rz+s)^2\, V_i(z) \ .\nonumber
\end{align}
Using these results, we obtain
\begin{align}
  V_1^\prime(z_1^\prime)V_2^\prime(z_2^\prime)V_3^\prime(z_3^\prime)
\,d  z_1^\prime\wedge d  z_2^\prime\wedge d  z_3^\prime
=
-\frac{2}{p}z_{12}z_{13}z_{23}
V_1(z_1)V_2(z_2)V_3(z_3)\,d  p\wedge d  q\wedge d  r \ .\nonumber
\end{align}
This reproduces the Faddeev-Popov procedure for the sphere amplitudes in
the Polyakov string, because the holomorphic part of an $SL(2,\mathbb{C})$ Haar measure 
\begin{align}
    \frac{1}{p}\,d  p\wedge d  q\wedge d  r
\end{align}
decouples with the Jacobian $z_{12}z_{13}z_{23}$
equal to the three-point function of a conformal $c$-ghost.

\subsection{Insertion of a non-marginal vertex operator in Polyakov string}
\label{non-marginal}

In section \ref{Evaluation of the String Entanglement Entropy} and
\ref{Generalization to Three-Dimensional Black Holes}, we evaluate an integrated $2s'$-point function of non-marginal vertex operators in the $\mathrm{LD}\times S^1$ CFT:
\begin{align}
     \frac{1}{\mathrm{vol}(SL(2,\mathbb{C}))}
\int d^2z'_1\cdots d^2z'_{s'}\,d^2w'_1\cdots d^2w'_{s'}
\left\langle 
W_+^{\prime \delta}(z'_1)\cdots W_+^{\prime \delta}(z'_{s'})\,
W_-^{\prime \delta}(w'_1)\cdots W_-^{\prime \delta}(w'_{s'})
\right\rangle
\end{align}
It suffices to keep only a single $W_+^{\prime\delta}$ or $W_-^{\prime\delta}$ inserted with the rest reduced to the marginal operator by setting $\delta=0$, because we are interested in the linear terms in $\delta$ expansions.
We first consider the case of a single $W_+^{\prime\delta}$ inserted at $z'=z'_3$. 
\begin{align}
     \frac{1}{\mathrm{vol}(SL(2,\mathbb{C}))}&
\int d^2z'_1\cdots d^2z'_{s'}\,d^2w'_1\cdots d^2w'_{s'}
\nonumber\\
&\times\left\langle 
W_+^{\prime}(z'_1)W_+^{\prime}(z'_2)W_+^{\prime \delta}(z'_{3}) W_+^{\prime}(z'_{4})\cdots W_+^{\prime}(z'_{s'})\,
W_-^{\prime}(w'_1)\cdots W_-^{\prime}(w'_{s'})
\right\rangle \ .
\label{singleW+}
\end{align}
Suppose the worldsheet coordinates $z'$  and $w'$ are related to $z_i$ and $w$ is $SL(2,\mathbb{C})$ as
\begin{align}
    z'=\frac{pz+q}{rz+s}\ ,~~~w'=\frac{pw+q}{rw+s} \ .
\end{align}
We pick up the three marginal operator $W'_+(z'_1),\,W'_+(z'_2)$ and $W'_-(w'_1)$ and apply the results obtained in the previous subsection.
We note that the non-marginal operator $W^{\prime\delta}_+(z'_3)$ makes it impossible  to factorize the $SL(2,\mathbb{C})$ volume from the integration of the location of the vertex operators. This is because the integrated non-marginal operator $W^{\prime\delta}_+(z'_3)\,d^2z'_3$ transforms as 
\begin{align}
   W^{\prime\delta}_+(z'_3)\,d^2z'_3
=\left|\frac{\partial z_3}{\partial z_3}\right|^{2(h_\delta-1)}W^{\delta}_+(z_3)\,d^2z_3
=\left|rz_3+s\right|^{2(h_\delta-1)}W^{\delta}_+(z_3)\,d^2z_3 \ ,
\label{transf:intW}
\end{align}
where $h_\delta$ is the conformal dimension of $W^{\prime\delta}_+$. Then, the $2s'$--point function is written as
\begin{align}
     \frac{1}{\mathrm{vol}(SL(2,\mathbb{C}))}&
\int 4\,d\mathrm{vol}(SL(2,\mathbb{C}))\, d^2z_3\cdots d^2z_{s'}\,d^2w_2\cdots d^2w_{s'}
\nonumber\\
&\times |z_1-z_2|^2\,|z_1-w_1|^2\,|z_2-w_1|^2\,|rz_3+s|^{2(h_\delta-1)}
\nonumber\\
&\times\left\langle 
W_+(z_{1})W_+(z_{2})W_+^{\delta}(z_{3})W_+(z_{4})\cdots W_+(z_{s'})\,
W_-(w_1)\cdots W_-(w_{s'})
\right\rangle \ .
\label{2spointW3}
\end{align}
Here,
\begin{align}
    d\mathrm{vol}(SL(2,\mathbb{C}))=\delta^2(ps-qr-1)\,d ^2p\,d ^2q\,d ^2r\,d ^2s \ ,
\end{align}
is the $SL(2,\mathbb{C})$ volume form.
We are thus led to evaluate 
\begin{align}
\frac{1}{\mathrm{vol}(SL(2,\mathbb{C}))}
\int d\mathrm{vol}(SL(2,\mathbb{C}))\, |rz+s|^{2\alpha}
=
\frac{1}{V}
\int d ^2r\,d ^2s\,\frac{1}{|s|^2} \,|rz+s|^{2\alpha} \ .
\label{int:sl2C}
\end{align}
with $\alpha$ being a constant and
\begin{align}
V=\int \frac{1}{|s|^2}d ^2r\,d ^2s \ .
\nonumber 
\end{align}
As this expression is ill-defined, 
we propose to regularize it as 
\begin{align}
  \hat{I}&:=\int d^2r\, d^2s\,\frac{1}{|s|^{2(1+\epsilon)}}\,
|rz+s|^{2\alpha}\,e^{-\rho_r|r|^2-\rho_s|s|^2} \ ,~~~
\nonumber\\
  \hat{V}&:=\int d^2r\, d^2s\, \frac{1}{|s|^{2(1+\epsilon)}}\,
|r|^{2\alpha}\,e^{-\rho_r|r|^2-\rho_s|s|^2} \ .
\nonumber
\end{align}
Here, $\epsilon,\rho_r,\rho_s$ are positive constants that are
sent to zero after the integration of $r$ and $s$.
It is easy to find
\begin{align}
  \hat{V}=\pi^2\Gamma(-\epsilon)\Gamma(1+\alpha)\frac{\rho_s^\epsilon}{\rho_r^{\alpha+1}}
  \ .\nonumber
\end{align}
We rewrite $\hat{I}$ as
\begin{align}
  \hat{I}=\int d^2s\,\frac{1}{|s|^{2(1+\epsilon)}}\,e^{-\rho_s|s|^2}
\int d^2r\,\frac{1}{\Gamma(-\alpha)}\int_0^\infty dt\,t^{-\alpha-1}\,e^{-t|rz+s|^2}\,e^{-\rho_r|r|^2}
\ .\nonumber
\end{align}
It is  straightforward to integrate $r$ and $s$:
\begin{align}
  \hat{I}=\frac{\pi^2\Gamma(-\epsilon)}{\Gamma(-\alpha)}
\int_0^\infty dt\, \frac{t^{-\alpha-1}}{t|z|^2+\rho_r}
\left(\rho_s+\frac{t\rho_r}{t|z|^2+\rho_r}\right)^\epsilon\ .\nonumber
\end{align}
Changing $t$ to an integration variable $x$ defined by 
\begin{align}
  x=\frac{\rho_r}{t|z|^2+\rho_r} \ ,\nonumber
\end{align}
we find
\begin{align}
  \hat{I}
&=
\frac{\pi^2\Gamma(-\epsilon)}{\rho_r\Gamma(-\alpha)}
\frac{|z|^{2\alpha}}{\rho_r^{\alpha}}\,
\left(\rho_s+\frac{\rho_r}{|z|^2}\right)^\epsilon\,
\int_0^1 dx\, x^{\alpha}\,(1-x)^{-\alpha-1}\,(1-ux)^\epsilon
\nonumber\\
&=
\pi^2\Gamma(-\epsilon)
\Gamma(1+\alpha)\,
\frac{|z|^{2\alpha}}{\rho_r^{\alpha+1}}
\left(\rho_s+\frac{\rho_r}{|z|^2}\right)^\epsilon\,
F(-\epsilon,1+\alpha,1;u)\ .\nonumber
\end{align}
Here,
\begin{align}
  u=\frac{\frac{\rho_r}{|z|^2}}{\rho_s+\frac{\rho_r}{|z|^2}} \ .\nonumber
\end{align}
It follows that
\begin{align}
  \frac{\hat{I}}{\hat{V}}
&
=
|z|^{2\alpha}
\,
(1-u)^{-\epsilon}\,F(-\epsilon,1+\alpha,1;u)
\nonumber\\
&=
|z|^{2\alpha}
\,
F(-\epsilon,-\alpha,1;u/(u-1))
\ ,\nonumber
\end{align}
with
\begin{align}
  \frac{u}{u-1}=-\frac{\rho_r}{\rho_s|z|^2} \ .\nonumber
\end{align}
Here, we used the formula
\begin{align}
  F(a,b,c;u)=(1-u)^{-a}\,F(a,c-b,c;u/(u-1)) \ .\nonumber
\end{align}
We obtain
\begin{align}
  \lim_{\epsilon\to 0}\frac{\hat{I}}{\hat{V}}
=|z|^{2\alpha}
F(0,\alpha,1;u/(u-1))
=|z|^{2\alpha}\ .\nonumber
\end{align}
with $\rho_r/\rho_s$ kept finite.

We also discuss the case with a single $W_-^{'\delta}$ inserted. In this case, it is useful to pick up $W_+^{'}(z'_1),W_-^{'}(z'_1)$
and $W_-^{'}(z'_2),$ and use the results in the previous subsection. The integrated non-marginal operator $W_-^{'\delta}(z')\,d^2z'$ transforms under the $SL(2,\mathbb{{C}})$ exactly in the same manner as in \eqref{transf:intW}. Then, we are again led to compute the integral \eqref{int:sl2C}. The resultant $2s'$-point correlation function in the $\mathrm{LD}\times S^1$ CFT can be computed by using the OPEs \eqref{ope-delta}, which show no distinction of the OPE between $W_+$ and $W_-$. Therefore, the integrated correlation function with a single insertion of $W_-^{'\delta}$ is identical to that of $W_+^{'\delta}$.

When computing \eqref{singleW+}, we fix the position of the three marginal operators $W'_+(z'_1),W'_+(z'_2)$ and $W'_-(w'_1)$ while that of $W_+^{'\delta}(z'_3)$ is to be integrated. We may fix the position of $W_+^{'\delta}(z'_3)$ instead of that of $W'_+$. Although the resultant $2s'$-point function takes a different form compared with  \eqref{2spointW3},  we note that the difference appears at the linear term in $\delta$ expansions. It then follows that the string entanglement entropy is independent of how to fix the position of three vertex operators. This is because the leading contribution to the integral over the position of $W_-$s is given by a term of $\mathcal{O}(\delta^2)$, see \eqref{exp:1/gamma}.

\bibliography{black_hole_entropy_from_string_entanglement}

\providecommand{\href}[2]{#2}\begingroup\raggedright\begin{thebibliography}{10}

\bibitem{Halder:2024gwe}
I.~Halder and D.~L. Jafferis, ``{Stretched horizon, replica trick and off-shell winding condensate, and all that},'' \href{https://arxiv.org/abs/2402.00932}{{\ttfamily arXiv:2402.00932 [hep-th]}}.

\bibitem{Sorkin:1984kjy}
R.~D. Sorkin, ``{1983 paper on entanglement entropy: ''On the Entropy of the Vacuum outside a Horizon''},'' in {\em {10th International Conference on General Relativity and Gravitation}}, vol.~2, pp.~734--736.
\newblock 1984.
\newblock \href{https://arxiv.org/abs/1402.3589}{{\ttfamily arXiv:1402.3589 [gr-qc]}}.

\bibitem{Bombelli:1986rw}
L.~Bombelli, R.~K. Koul, J.~Lee, and R.~D. Sorkin, ``{A Quantum Source of Entropy for Black Holes},'' \href{https://dx.doi.org/10.1103/PhysRevD.34.373}{{\em Phys. Rev. D} {\bfseries 34} (1986) 373--383}.

\bibitem{Witten:2024upt}
E.~Witten, ``{Introduction to Black Hole Thermodynamics},'' \href{https://arxiv.org/abs/2412.16795}{{\ttfamily arXiv:2412.16795 [hep-th]}}.

\bibitem{Maldacena:2001kr}
J.~M. Maldacena, ``{Eternal black holes in anti-de Sitter},'' \href{https://dx.doi.org/10.1088/1126-6708/2003/04/021}{{\em JHEP} {\bfseries 04} (2003) 021}, \href{https://arxiv.org/abs/hep-th/0106112}{{\ttfamily arXiv:hep-th/0106112}}.

\bibitem{Susskind:1993ws}
L.~Susskind, ``{Some speculations about black hole entropy in string theory},'' \href{https://arxiv.org/abs/hep-th/9309145}{{\ttfamily arXiv:hep-th/9309145}}.

\bibitem{Susskind:1994sm}
L.~Susskind and J.~Uglum, ``{Black hole entropy in canonical quantum gravity and superstring theory},'' \href{https://dx.doi.org/10.1103/PhysRevD.50.2700}{{\em Phys. Rev. D} {\bfseries 50} (1994) 2700--2711}, \href{https://arxiv.org/abs/hep-th/9401070}{{\ttfamily arXiv:hep-th/9401070}}.

\bibitem{Jafferis:2021ywg}
D.~L. Jafferis and E.~Schneider, ``{Stringy ER = EPR},'' \href{https://dx.doi.org/10.1007/JHEP10(2022)195}{{\em JHEP} {\bfseries 10} (2022) 195}, \href{https://arxiv.org/abs/2104.07233}{{\ttfamily arXiv:2104.07233 [hep-th]}}.

\bibitem{FateevZamolodchikovUnpublished}
V.~Fateev, A.~Zamolodchikov, and A.~Zamolodchikov, ``Unpublished.''
\newblock Unpublished manuscript.

\bibitem{Kazakov:2000pm}
V.~Kazakov, I.~K. Kostov, and D.~Kutasov, ``{A Matrix model for the two-dimensional black hole},'' \href{https://dx.doi.org/10.1016/S0550-3213(01)00606-X}{{\em Nucl. Phys. B} {\bfseries 622} (2002) 141--188}, \href{https://arxiv.org/abs/hep-th/0101011}{{\ttfamily arXiv:hep-th/0101011}}.

\bibitem{Witten:1991yr}
E.~Witten, ``{On string theory and black holes},'' \href{https://dx.doi.org/10.1103/PhysRevD.44.314}{{\em Phys. Rev. D} {\bfseries 44} (1991) 314--324}.

\bibitem{Itzhaki:2018glf}
N.~Itzhaki, ``{Stringy instability inside the black hole},'' \href{https://dx.doi.org/10.1007/JHEP10(2018)145}{{\em JHEP} {\bfseries 10} (2018) 145}, \href{https://arxiv.org/abs/1808.02259}{{\ttfamily arXiv:1808.02259 [hep-th]}}.

\bibitem{Maldacena:2013xja}
J.~Maldacena and L.~Susskind, ``{Cool horizons for entangled black holes},'' \href{https://dx.doi.org/10.1002/prop.201300020}{{\em Fortsch. Phys.} {\bfseries 61} (2013) 781--811}, \href{https://arxiv.org/abs/1306.0533}{{\ttfamily arXiv:1306.0533 [hep-th]}}.

\bibitem{Callan:1988hs}
C.~G. Callan, Jr., R.~C. Myers, and M.~J. Perry, ``{Black Holes in String Theory},'' \href{https://dx.doi.org/10.1016/0550-3213(89)90172-7}{{\em Nucl. Phys. B} {\bfseries 311} (1989) 673--698}.

\bibitem{Halder:2022ykw}
I.~Halder, D.~L. Jafferis, and D.~K. Kolchmeyer, ``{A duality in string theory on AdS$_{3}$},'' \href{https://dx.doi.org/10.1007/JHEP07(2023)049}{{\em JHEP} {\bfseries 07} (2023) 049}, \href{https://arxiv.org/abs/2208.00016}{{\ttfamily arXiv:2208.00016 [hep-th]}}.

\bibitem{VanRaamsdonk:2010pw}
M.~Van~Raamsdonk, ``{Building up spacetime with quantum entanglement},'' \href{https://dx.doi.org/10.1142/S0218271810018529}{{\em Gen. Rel. Grav.} {\bfseries 42} (2010) 2323--2329}, \href{https://arxiv.org/abs/1005.3035}{{\ttfamily arXiv:1005.3035 [hep-th]}}.

\bibitem{Hikida:2008pe}
Y.~Hikida and V.~Schomerus, ``{The FZZ-Duality Conjecture: A Proof},'' \href{https://dx.doi.org/10.1088/1126-6708/2009/03/095}{{\em JHEP} {\bfseries 03} (2009) 095}, \href{https://arxiv.org/abs/0805.3931}{{\ttfamily arXiv:0805.3931 [hep-th]}}.

\bibitem{Karczmarek:2004bw}
J.~L. Karczmarek, J.~M. Maldacena, and A.~Strominger, ``{Black hole non-formation in the matrix model},'' \href{https://dx.doi.org/10.1088/1126-6708/2006/01/039}{{\em JHEP} {\bfseries 01} (2006) 039}, \href{https://arxiv.org/abs/hep-th/0411174}{{\ttfamily arXiv:hep-th/0411174}}.

\bibitem{Maldacena:2005hi}
J.~M. Maldacena, ``{Long strings in two dimensional string theory and non-singlets in the matrix model},'' \href{https://dx.doi.org/10.1088/1126-6708/2005/09/078}{{\em JHEP} {\bfseries 09} (2005) 078}, \href{https://arxiv.org/abs/hep-th/0503112}{{\ttfamily arXiv:hep-th/0503112}}.

\bibitem{Chen:2021emg}
Y.~Chen and J.~Maldacena, ``{String scale black holes at large D},'' \href{https://dx.doi.org/10.1007/JHEP01(2022)095}{{\em JHEP} {\bfseries 01} (2022) 095}, \href{https://arxiv.org/abs/2106.02169}{{\ttfamily arXiv:2106.02169 [hep-th]}}.

\bibitem{Halder:2023adw}
I.~Halder and D.~L. Jafferis, ``{Thermal Bekenstein-Hawking entropy from the worldsheet},'' \href{https://dx.doi.org/10.1007/JHEP05(2024)136}{{\em JHEP} {\bfseries 05} (2024) 136}, \href{https://arxiv.org/abs/2310.02313}{{\ttfamily arXiv:2310.02313 [hep-th]}}.

\bibitem{Agia:2022srj}
N.~Agia and D.~L. Jafferis, ``{Angular Quantization in CFT},'' \href{https://arxiv.org/abs/2204.11872}{{\ttfamily arXiv:2204.11872 [hep-th]}}.

\bibitem{Gibbons:1976ue}
G.~W. Gibbons and S.~W. Hawking, ``{Action Integrals and Partition Functions in Quantum Gravity},'' \href{https://dx.doi.org/10.1103/PhysRevD.15.2752}{{\em Phys. Rev. D} {\bfseries 15} (1977) 2752--2756}.

\bibitem{Lewkowycz:2013nqa}
A.~Lewkowycz and J.~Maldacena, ``{Generalized gravitational entropy},'' \href{https://dx.doi.org/10.1007/JHEP08(2013)090}{{\em JHEP} {\bfseries 08} (2013) 090}, \href{https://arxiv.org/abs/1304.4926}{{\ttfamily arXiv:1304.4926 [hep-th]}}.

\bibitem{Dorn:1994xn}
H.~Dorn and H.~J. Otto, ``{Two and three point functions in Liouville theory},'' \href{https://dx.doi.org/10.1016/0550-3213(94)00352-1}{{\em Nucl. Phys. B} {\bfseries 429} (1994) 375--388}, \href{https://arxiv.org/abs/hep-th/9403141}{{\ttfamily arXiv:hep-th/9403141}}.

\bibitem{Zamolodchikov:1995aa}
A.~B. Zamolodchikov and A.~B. Zamolodchikov, ``{Structure constants and conformal bootstrap in Liouville field theory},'' \href{https://dx.doi.org/10.1016/0550-3213(96)00351-3}{{\em Nucl. Phys. B} {\bfseries 477} (1996) 577--605}, \href{https://arxiv.org/abs/hep-th/9506136}{{\ttfamily arXiv:hep-th/9506136}}.

\bibitem{Fischler:1986tb}
W.~Fischler and L.~Susskind, ``{Dilaton Tadpoles, String Condensates and Scale Invariance. 2.},'' \href{https://dx.doi.org/10.1016/0370-2693(86)90514-9}{{\em Phys. Lett. B} {\bfseries 173} (1986) 262--264}.

\bibitem{Fischler:1986ci}
W.~Fischler and L.~Susskind, ``{Dilaton Tadpoles, String Condensates and Scale Invariance},'' \href{https://dx.doi.org/10.1016/0370-2693(86)91425-5}{{\em Phys. Lett. B} {\bfseries 171} (1986) 383--389}.

\bibitem{Ahmadain:2022eso}
A.~Ahmadain and A.~C. Wall, ``{Off-shell strings II: Black hole entropy},'' \href{https://dx.doi.org/10.21468/SciPostPhys.17.1.006}{{\em SciPost Phys.} {\bfseries 17} no.~1, (2024) 006}, \href{https://arxiv.org/abs/2211.16448}{{\ttfamily arXiv:2211.16448 [hep-th]}}.

\end{thebibliography}\endgroup
\bibliographystyle{utphys}
\end{document}